\documentclass[aps,twocolumn,showpacs,preprintnumbers,nofootinbib,prd,superscriptaddress,floatfix,10pt]{revtex4-2}

\usepackage{graphicx,amssymb,amsmath,amsthm,amsfonts}
\usepackage{dcolumn}
\usepackage{bm}
\usepackage{subcaption}
\usepackage{float}
\usepackage{color}

\begin{document}


\title{Onset of spontaneous scalarization in dark matter stars}


\author{Junya Tanaka}
\email{jtlightc555@gmail.com}
\affiliation{Independent Researcher, Tokyo, Japan}


\date{\today}
\begin{abstract}
We investigated the onset of spontaneous scalarization in dark matter stars. In scalar–tensor theories, the coupling between matter and a scalar field generates an effective mass, which is known to trigger the growth of the scalar field around stars and black holes (spontaneous scalarization). On the other hand, models in which the scalar field couples not only to baryonic matter but also to dark matter have been discussed in the literature. Moreover, in models with self-interacting dark matter, it is known that dark matter can form compact objects, dark stars (DS). Within the framework of scalar–tensor theories, we studied the parameter ranges in which the growth of the scalar field occurs in  dark stars. As a result, we found that the tachyonic instability associated with spontaneous scalarization occurs in some parameter regions with a negative coupling parameter $\beta_d$.  We also investigated double-degenerate stars (DDSs), which consist of baryonic matter and dark matter, and found that the baryonic component weakens the tachyonic instability. As a result, dark stars containing baryons can still exhibit the tachyonic instability, whereas neutron stars containing a dark matter component do not develop tachyonic instability induced by the dark matter.
\end{abstract}


\maketitle
\section{\label{intro}Introduction}
Scalar–tensor theory is an extension of general relativity (GR) in which scalar degrees of freedom are added to the gravitational sector. Its general action is written in the Jordan frame as follows \cite{Fujii:2003pa}:
\begin{align}\label{Jordanframe}
S=&\frac{1}{16\pi G}\int d^4x \sqrt{-g}(F(\phi)R-Z(\phi)g^{\mu\nu}\partial_{\mu}\phi \partial_{\nu}\phi \nonumber \\
&-U(\phi))+S(\Psi_m;g_{\mu \nu}))
\end{align}
Here $R$ is the Ricci scalar, $g_{\mu\nu}$ is the spacetime metric, $\phi$ is the scalar field, and $G$ is a constant.
By specifying $F, Z,$ and $U$, one can recover individual theories.
(For example, for Brans–Dicke theory, $F=\phi$, $Z=\omega/\phi$, and $U=0$ \cite{PhysRev.124.925}.)
The scalar field is nonminimally coupled to gravity and thus couples indirectly to matter.\par

On the other hand, the scalar field is also suggested to be the source of dark energy, which constitutes more than 70 percent of the energy content of the Universe. This is modeled as quintessence \cite{Ratra:1987rm}. If the scalar field plays a cosmological role in the evolution of the Universe, its mass must be extremely small or nearly zero (of order $H_0$, the Hubble constant \cite{2004PhRvD..69d4026K}). However, such a scalar field is ruled out by equivalence principle experiments and constraints on fifth forces on Earth. This implies that the scalar field must be massive in terrestrial environments. To resolve this apparent contradiction, various screening mechanisms for scalar fields, such as the chameleon mechanism \cite{2004PhRvD..69d4026K,2004PhRvL..93q1104K}, have been proposed.\par

However, these observational constraints mainly apply to scalarization associated with baryonic matter. In the standard cosmological picture, dark matter plays a crucial role in the formation of galaxies and large-scale structure of the Universe \cite{1982ApJ...263L...1P,1984Natur.311..517B,1985ApJ...292..371D,1978MNRAS.183..341W,1997ApJ...490..493N,2012AnP...524..507F}. If dark matter is composed of particle-like matter, it may interact with a scalar field. In that case, the coupling to dark matter may evade the observational constraints mentioned above. Models of interactions between dark matter and dark energy have been proposed previously, and their effects have been studied cosmologically from the viewpoints of quintessence and scalar–tensor theories \cite{PhysRevLett.64.123,2000PhRvD..62d3511A,2004ApJ...604....1F,2013PhRvD..88f3005F,2014PhRvL.113r1301S}. Introducing such interactions is rather reasonable from the perspective of the coincidence problem \cite{1999PhRvL..82..896Z,2006IJMPD..15.1753C}.\par
There are also studies on trends outside of cosmology.
Dark matter stars coupled to scalar fields have been studied primarily in the context of chameleon dark matter stars \cite{2013PhRvD..88f3005F}. In this model, the scalar field acts as an additional force, deforming stellar structure and the mass-radius relationship. In these models, dark sector interactions deform the equilibrium state but do not induce dynamic instabilities.\par

When the scalar field couples to the trace of the energy--momentum tensor of matter or to curvature invariants, an effective mass is generated for the scalar field. If the square of this effective mass becomes negative, the scalar field develops a tachyonic instability. At the linear level, this instability is described by exponentially growing scalar-field perturbations. The nonlinear evolution of this instability may lead to spontaneous scalarization, in which the scalar field acquires a nontrivial configuration around compact objects such as neutron stars and black holes. Such an instability occurs only in strong-field regions where the scalar field couples sufficiently strongly to matter or curvature. The possibility of spontaneous scalarization was first proposed for neutron stars by Damour and Esposito-Farèse \cite{PhysRevLett.70.2220,1996PhRvD..54.1474D}. Since then, scalarization of neutron stars has been studied extensively, and the original DEF model has been found to be strongly constrained by pulsar observations \cite{2024RvMP...96a5004D}. However, the tachyonic instability in compact objects composed of dark matter remains largely unexplored.\par

Self-interacting dark matter was proposed to resolve the tensions between collisionless cold dark matter and astrophysical observations \cite{2000PhRvL..84.3760S,2013MNRAS.430...81R,2018PhR...730....1T} (such as the cuspy core problem, the satellite galaxy problem, and the “too big to fail” problem). In the early Universe, asymmetric dark matter (ADM) \cite{2009PhRvD..79k5016K,2013IJMPA..2830028P,2014PhR...537...91Z} with self-interactions may form compact astrophysical objects \cite{2015PhRvD..92f3526K}. These objects are sometimes discussed as alternatives to neutron stars or black holes. If dark matter interacts with a scalar field, scalarization may occur in such dark matter stars. In that case, the dark matter stars themselves, together with the detection of scalar fields, could provide clues to forming a physical picture of the dark sector.\par
Dark stars can be broadly classified into two categories: those composed entirely of dark matter and those consisting of both dark matter and baryonic gas. The latter can be further divided into two types. In one type, baryonic gas forms a central core surrounded by a dark matter envelope, whereas in the other, dark matter accumulates at the center of a baryonic star, such as a neutron star, forming a dark matter core \cite{PhysRevD.84.107301}. From an astrophysical perspective, the former configuration is expected to arise naturally because the gravitational potential of a dark star can capture surrounding baryonic gas or accrete matter from a companion star, leading to the formation of a mixed baryon--dark matter core at its center. The latter configuration corresponds to the accumulation of dark matter inside a neutron star due to its strong gravitational field. These objects are collectively referred to as doubly degenerate stars (DDSs) \cite{Luo_Xin-Lian_2008}.
If the objective is to investigate the intrinsic properties of dark matter stars, pure dark stars provide the most appropriate model. However, in realistic astrophysical environments, both pure dark stars and mixed configurations are expected to exist. Therefore, in this paper, we consider both pure dark stars and mixed dark matter--baryon stars.\par
Previous studies of scalarization have mainly focused on scalar couplings to baryonic matter. However, it is unclear whether the same tachyonic instability would still arise if dark matter were to interact with the scalar field. No systematic investigation of this issue has been conducted to date. Given that the scalarization of neutron stars has been effectively ruled out, investigating the scalarization of dark stars is considered significant for both the verification of scalar tensor theory and the development of theories of dark matter.
\par
Based on the above considerations, we investigate the linear onset of spontaneous scalarization in pure dark matter stars and DDSs by analyzing the tachyonic instability of scalar-field perturbations. The content of this paper is as follows. In Sec.~\ref{stt}, we describe the basic formulation of scalar–tensor theories and the perturbation equations. In Sec.~\ref{DM}, we explain the self-interacting dark matter model and its parameter constraints. Section~\ref{BC} is devoted to the boundary conditions for the perturbation equations. Numerical results are presented in Sec.~\ref{NR}, and conclusions are given in Sec.~\ref{con}. Throughout this paper, we use units with $c=G=1$.

\section{\label{stt}The scalar-tensor theory}
The general action of scalar–tensor theory in the Einstein frame is obtained by the conformal transformation
\begin{equation}
g_{\mu\nu} = A^2(\phi) g^*{\mu\nu}
\end{equation}
and is written as \cite{T_Damour_1992}
\begin{align}
I &= \frac{1}{16\pi G_*}\int d^4x \sqrt{-g_*}(R_* - 2 g_*^{\mu\nu}\varphi_{,\mu}\varphi_{,\nu})   \notag \\
&+ I_m[\Psi_m, A^2(\varphi) g^*_{\mu\nu}] .
\end{align}

Here, $G_*$ is the gravitational constant. $\varphi$ is the scalar field in the Einstein frame. The asterisk $*$ indicates quantities defined in the Einstein frame; $g^*_{\mu\nu}$ is the metric in the Einstein frame, and $R*$ is the Ricci scalar. The function $A(\varphi)$ appearing in the matter action $I_m$ is the coupling function that describes the interaction between the scalar field and matter. The field equations are given by

\begin{equation}
G^*_{\mu\nu} =8\pi G_* T_{*\mu\nu}+ 2\left(\varphi_{,\mu}\varphi_{,\nu}-\frac12 g^*_{\mu\nu} g_*^{\alpha\beta}\varphi_{,\alpha}\varphi_{,\beta}\right)
\end{equation}
\begin{equation}
\Box_* \varphi = -4\pi G_* \alpha(\varphi) T_*
\end{equation}
where
\begin{equation}
T^{\mu\nu}_*\equiv\frac{2}{\sqrt{-g_*}}\frac{\delta I_m[\Psi_m, A^2(\varphi) g^*_{\mu\nu}]}{\delta g^*_{\mu\nu}}= A^6(\varphi) T^{\mu\nu}
\end{equation}

\begin{equation}
T_* \equiv T^\mu_{*\mu} = T^{\mu\nu}_* g^*_{\mu\nu}
\end{equation}

Several models for interactions between dark matter and a scalar field (dark energy) have been proposed. For example, there are models of the Damour type \cite{PhysRevLett.64.123} with an $S_m[\psi_m,e^{2\beta \varphi}g_{\mu\nu}]$ coupling, as well as models such as those proposed by Amendola \cite{2000PhRvD..62d3511A}, in which energy transfer terms between dark matter and dark energy appear in the matter continuity equations:
\begin{align}
\dot{\rho_{\text{DM}}}+3H\rho_{\text{DM}}=Q \\
\dot{\rho_{\text{DE}}}+3H(1+\omega_{\text{DE}})\rho_{\text{DE}}=-Q
\end{align}
Here, $\rho_{\text{DM}}$, $\rho_{\text{DE}}$ is the energy density of dark matter and dark energy, $\omega_{\text{DE}}$ is equation of state of dark energy, $H$ is the Hubble parameter. There are various types of $Q$. While $Q\propto \dot{\varphi}$ corresponds to the Damour type, there is a phenomenological type where $Q=\xi H \rho_{\text{DM}}$. These models have been constrained mainly from a cosmological perspective. For the former, the constraint is $\beta \le O(1)$, while for the latter there are also limitations, $\xi \le O(0.01)$.
Motivated by this, we model the coupling function as a linear form,
\begin{equation}
\alpha(\varphi)=\beta_d \varphi .
\end{equation}
This form is the same as that commonly assumed for the coupling to ordinary baryonic matter (DEF model), namely,
\begin{equation}
\alpha(\varphi)=\alpha_0+\beta \varphi ,
\end{equation}
where
\begin{equation}
\alpha_0 \simeq 0 .
\end{equation}
For the sake of comparison, we denote the dimensionless constant in the coupling function between dark matter and the scalar field as $\beta_d$, distinguishing it from the dimensionless constant $\beta$ in the coupling between baryonic matter and the scalar field.
This coupling form represents a minimal phenomenological setup, allowing us to exclude effects such as energy exchange between DM and DE or informal coupling, and enabling us to estimate the pure contribution from the dark star's profile. This choice also has the advantage that it allows for a straightforward comparison with previous results for couplings to baryonic matter.

If the coupling is linear at the field level, the solutions of the field equations admit a constant scalar field and background solutions identical to those of GR. Consequently, the background solutions in the Einstein frame and the Jordan frame coincide with the GR solutions \cite{1997PThPh..98..359H}:
\begin{align}
g_{\mu\nu}^*& = g_{\mu\nu} = g_{\mu\nu}^{(GR)} ,\\
T_{\mu\nu}^*& = T_{\mu\nu} = T_{\mu\nu}^{(GR)} ,\\
\varphi& = \varphi_0 .
\end{align}
Therefore, from this point on, even for physical quantities within the Einstein frame, the * symbol will be omitted.
Considering linear perturbations of the scalar field,
\begin{equation}
\varphi = \varphi_0 + \delta\varphi ,
\end{equation}
the variation of the coupling function $A(\varphi)$ is of order $(\delta\varphi)^2$, so that $g^*_{\mu\nu}$ and $T^*_{\mu\nu}$ receive changes only at second order or higher. At first order, the metric and matter perturbations decouple from the scalar perturbation. Therefore, the field equations can be written as
\begin{align}
\delta G*{\mu\nu} &= 8\pi \delta T_{\mu\nu} ,\\
\Box^{(GR)} \delta\varphi &= -4\pi \beta_d T^{(GR)} \delta\varphi \label{scalareq} .
\end{align}
Here, $\delta T_{\mu\nu}$ denotes the first-order variation with respect to $T_{\mu\nu}^{(GR)}$.

The general metric of a spherically symmetric and static spacetime is given by
\begin{equation}\label{eq:metric}
ds^2 = -e^{2\Phi(r)} dt^2 + e^{2\Psi(r)} dr^2
+ r^2(d\theta^2 + \sin^2\theta\, d\phi^2)
\end{equation}
 In this spacetime, the field equation (\ref{scalareq}) becomes
\begin{align}\label{symmetricspacetime}
&-e^{-2\Phi} \frac{\partial^2 \delta\varphi}{\partial t^2}
+ \frac{e^{-\Phi-\Psi}}{r^2}
\frac{\partial}{\partial r}
\left(e^{\Phi-\Psi} r^2 \frac{\partial \delta\varphi}{\partial r}\right)\\\notag
&+ \frac{1}{r^2} \Big[\frac{1}{\sin\theta}\frac{\partial}{\partial \theta}\Big(\sin\theta\frac{\partial\delta\varphi}{\partial \theta}\Big)+\frac{1}{\sin^2\theta}\frac{\partial^2\delta\varphi}{\partial\phi^2}\Big]
+ 4\pi \beta_d T \delta\varphi = 0
\end{align}

Decomposing the perturbation variable as
\begin{equation}
\delta\varphi
= e^{i\omega t}
\sum_{l,m} \frac{\psi_{\omega lm}(r)}{r} Y_l^m(\theta,\phi)
\end{equation}
the field equation is finally reduced to a Schrödinger-type equation:
\begin{equation}\label{schrodinger}
\frac{d^2\psi_{\omega lm}}{dr_*^2}
+ [\omega^2 - V(r_*)]\psi_{\omega lm} = 0
\end{equation}
  Here, $r_*$ is the generalized tortoise coordinate, defined by
\begin{equation}
dr_* = e^{\Psi-\Phi} dr
\end{equation}
Furthermore, since this paper is limited to the case where $\ell=0$ ($m=0$), the $\ell$ and $m$ subscripts of $\psi$ will be omitted hereafter.
The function $V(r_*)$ is the effective potential, given by

\begin{equation}\label{eq:Veff}
V(r_*) =
\frac{\Phi'-\Psi'}{r} e^{2(\Phi-\Psi)}
+ \frac{l(l+1)}{r^2} e^{2\Phi}
- 4\pi \beta_d T e^{2\Phi}
\end{equation}
The last term corresponds to the effective mass induced by the coupling between the scalar field and dark matter.

\begin{figure*}[th]
\centering
\begin{minipage}{1.0\columnwidth}
\centering
\includegraphics[width=\columnwidth]{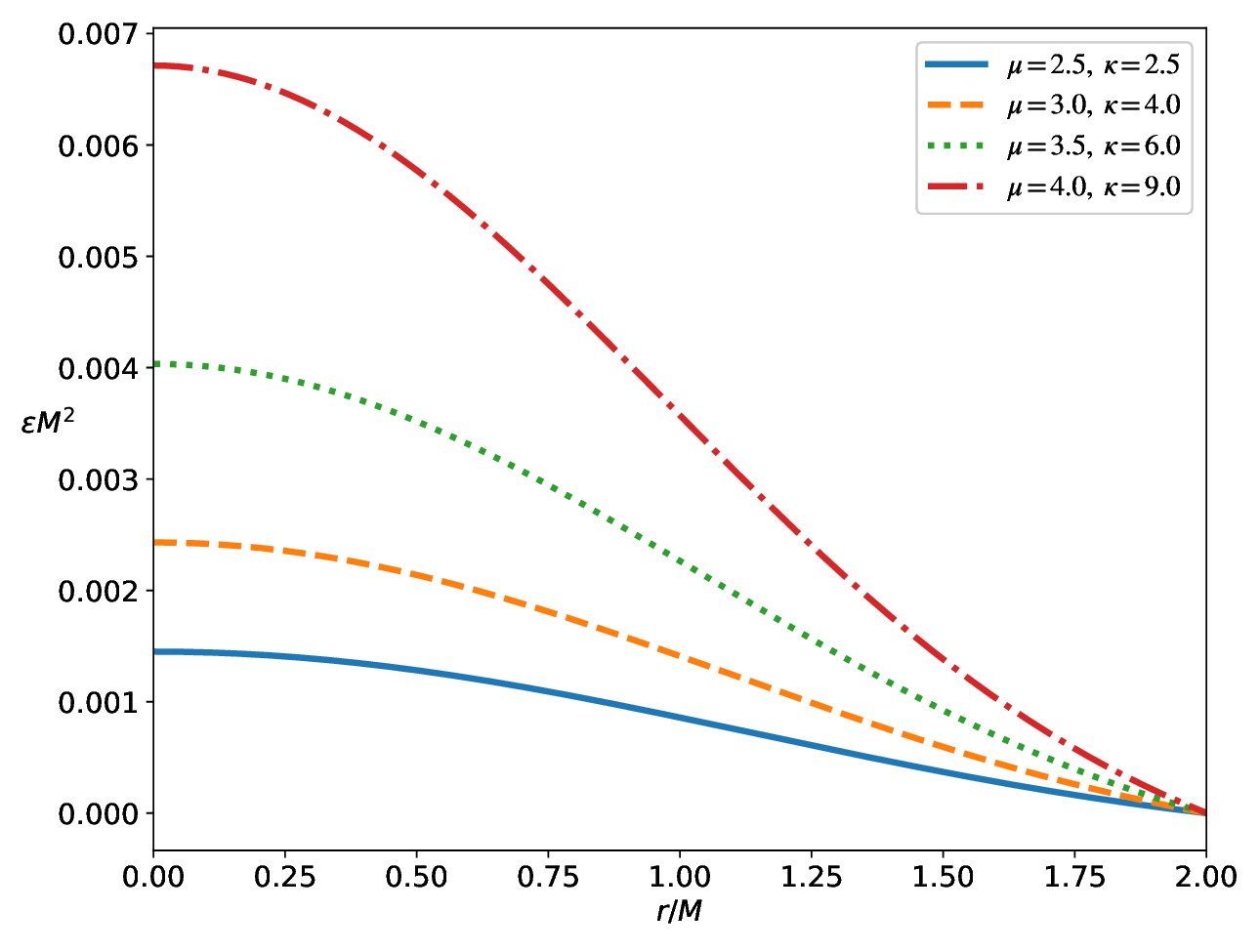}
\caption{\label{profile2}Energy density profiles for each $(\mu,\kappa)$. 
The stellar mass is fixed at $M=2.0M_{\odot}$. 
The horizontal 
axis denotes the dimensionless radius $r/M$.}
\end{minipage}
\begin{minipage}{1.0\columnwidth}
\centering
\includegraphics[width=\columnwidth]{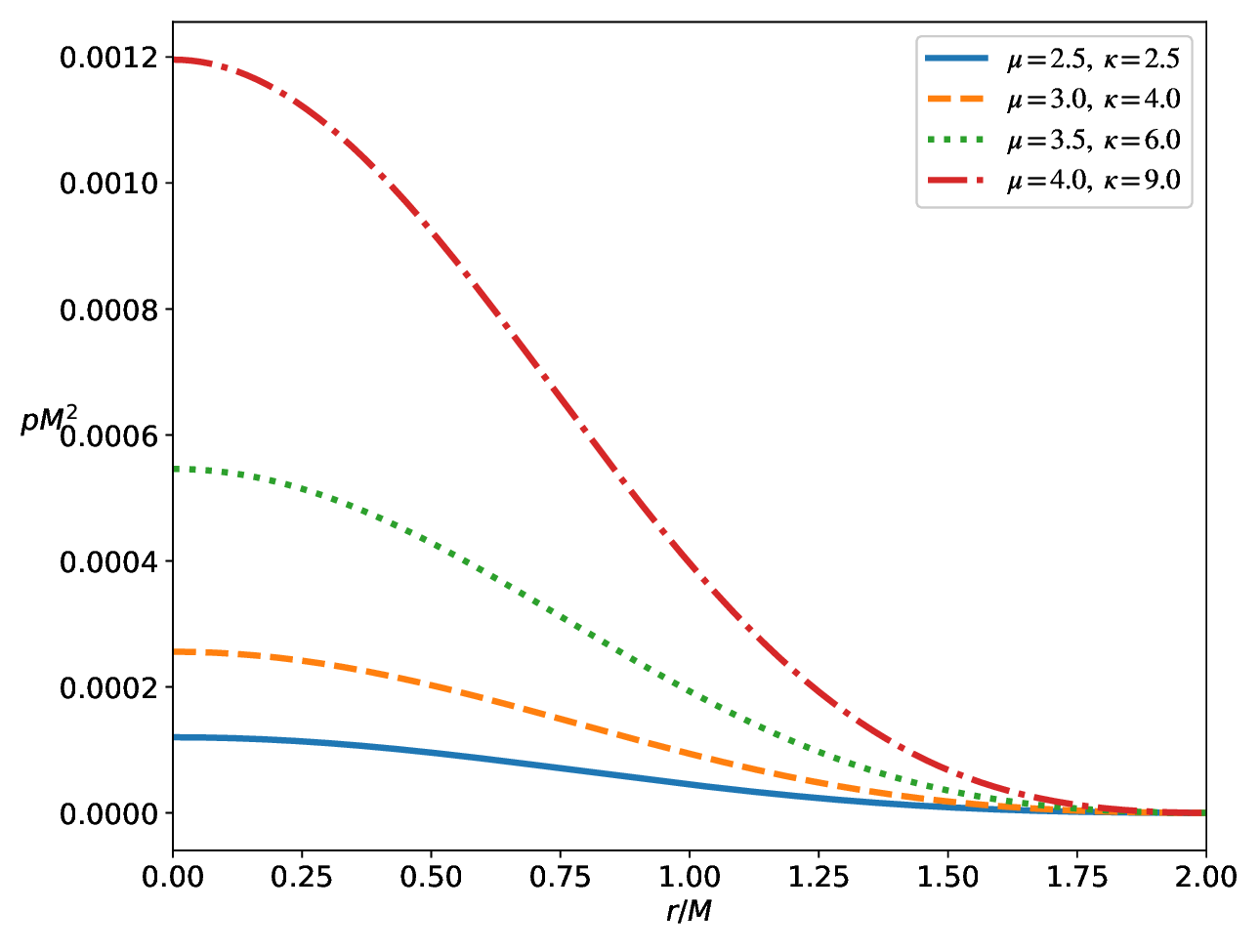}
\caption{\label{profile}Pressure profiles of stars at each $(\mu,\kappa)$. 
The stellar mass
 is fixed at $M=2.0M_{\odot}$. 
The horizontal axis denotes the dimensionless radius $r/M$.}
\end{minipage}
\end{figure*}

\section{\label{DM}Dark star models}
As a model of dark matter, we consider self-interacting asymmetric fermionic dark matter. The self-interaction is mediated by Yukawa-type interactions, and both an attractive type  and a repulsive type  can be considered. In general, while the typical masses of compact objects such as neutron stars and black holes satisfy $M > M_{\odot}$, the attractive type tends to produce objects with masses that are several orders of magnitude smaller. Therefore, in this study, we focus only on the repulsive type. Compact objects are formed by this self-interacting dark matter. In this section, we will first discuss stars composed entirely of dark matter. We assume that the spacetime of the star is described by a static and spherically symmetric spacetime (\ref{symmetricspacetime}). The energy–momentum tensor is taken to be that of a perfect fluid:
\begin{equation}
T^{\mu\nu} = (\epsilon+p)u^\mu u^\nu + pg^{\mu\nu}.
\end{equation}
Here, $\epsilon$ is the energy density of dark matter, and $p$ is the pressure. Equations of state (EOS) for dark matter, where $m_X$ is the mass of dark matter, are \cite{2015PhRvD..92f3526K}
\begin{align}
\epsilon &= \frac{m_{_X}^4}{\hbar^3}\left[\xi(x)+\frac{2}{9\pi^3}\frac{\alpha_{_X}}{\hbar}\frac{m_{_X}^2}{m_\phi^2}x^6\right] ,\nonumber\\
\xi(x) &= \frac{1}{8\pi^2}\left[x\sqrt{1+x^2}(2x^2+1)-\ln(x+\sqrt{1+x^2})\right],\label{eq: DMeos1}
\end{align}
and
\begin{align}
p &= \frac{m_{_X}^4}{\hbar^3}\left[\chi(x)+\frac{2}{9\pi^3}\frac{\alpha_{_X}}{\hbar}\frac{m_{_X}^2}{m_\phi^2}x^6\right] ,\nonumber\\
\chi(x) &= \frac{1}{8\pi^2}\left[x\sqrt{1+x^2}\left(\frac{2}{3}x^2-1\right)+\ln(x+\sqrt{1+x^2})\right]\label{eq: DMeos2},
\end{align}
Here $x=p_F/m_X$ is dimensionless Fermi momentum of dark matter fermion. For our numerical computation, we introduce a parameter $v=x^2$ related to the central Fermi momentum through
\begin{equation}\label{eq:vc}
x_c=\sqrt{|v_c|}
\end{equation}
where $x_c$ denotes the value of $x$ at the stellar center. Since the equilibrium sequence is uniquely determined by the central density (or equivalently by $x_c$), we use $v_c$ as a parameter labeling the stellar models.\par
The self-interaction between dark matter particles is mediated by a boson with rest mass $m_{\phi}$. The coupling constant $\alpha_{_X}$ characterizes the strength of this interaction. In the case of a repulsive interaction is $\alpha_{_X}>0$.\par

The equilibrium structure of a dark matter star is determined by the above equation of state (EOS), together with the equation for the metric function,
\begin{equation}
\frac{d\Phi}{dr} =  \frac{m+4\pi pr^3}{r(r-2m)},
\label{eq: dnudr}
\end{equation}
and the Tolman–Oppenheimer–Volkoff (TOV) equation for hydrostatic balance,
\begin{equation}
\frac{dp}{dr} =  -\frac{(\epsilon+p)(m+4\pi pr^3)}{r(r-2m)},
\label{eq: dpidr}
\end{equation}
by integration. Here, $m(r)=\int_0^r \epsilon 4\pi r'^2 dr'$.

Self-interacting dark matter was proposed in order to resolve the problems that cold dark matter faces in galaxy formation. Dark matter that resolves these problems while remaining consistent with existing galaxy formation scenarios is subject to the following constraint on the self-interaction cross section \cite{2004ApJ...606..819M,2008ApJ...679.1173R,2016PhRvL.116d1302K}:
\begin{equation}
	\sigma = 5\times 10^{-23}({\rm cm}^2)
	\left(\frac{\alpha_{_X}}{10^{-2}}\right)^2
	\left(\frac{m_{_X}}{10 {\rm GeV}}\right)^2
	\left(\frac{m_\phi}{10{\rm MeV}}\right)^{-4},
	\label{eq: Zurek-crossection}
\end{equation}
\begin{equation}
0.1 \le \frac{\sigma}{m_{_X}} \le 1\quad({\rm cm}^2{\rm g}^{-1}),
\label{eq: Kouvaris-Nielsen-constraint}
\end{equation}
which can be rewritten as follows:
\begin{align}
	0.36\le 
	\left(\frac{\alpha_{_X}}{10^{-2}}\right)^2
	\left(\frac{m_{_X}}{100 m_\phi}\right)^4
	\left(\frac{m_{_X}}{1{\rm GeV}}\right)^{-3} 
	\le 3.6. 
	\label{eq: constraint1}
\end{align}
Here, the two parameters $\mu$ and $\kappa$ are defined as
\begin{equation}
	\mu \equiv \left(\frac{m_{_X}}{1 {\rm GeV}}\right),
	\label{eq: def-mu}
\end{equation}
and
\begin{equation}
	\kappa \equiv 
	\left(\frac{m_{_X}}{100m_\phi}\right)^2\frac{\alpha_{_X}}{10^{-2}}.
	\label{eq: def-kappa2},
\end{equation}
then the constraint (\ref{eq: constraint1}) is rewritten as
\begin{equation}
	0.36 \le \mu^{-3}\kappa^2 \le 3.6 .
	\label{eq: constraint2}.
\end{equation}

\section{\label{BC}Boundary conditions}

The general solution of eq.~(\ref{schrodinger}) can be written as
\begin{equation}
\psi(r_*) = C_+(\omega)\psi_+(r_*) + C_-(\omega)\psi_-(r_*)
\end{equation}
as a linear combination of purely ingoing waves $\psi_+$ and purely outgoing waves $\psi_-$. Here, $C_+$ and $C_-$ are coefficients. The boundary conditions are such that at infinity, $\psi_+ \to e^{i\omega r_*}$ and $\psi_- \to e^{-i\omega r_*}$. For practical calculations, to avoid irregular points at infinity, we impose the boundary conditions at a finite radius in the asymptotic region $|\omega| r \gg 1$ and $r \gg M$, the solutions take the form
\begin{align}
\psi_+(r_*)&=e^{i\omega r_*}\sum_{\nu=0}^{\infty} b^+_{\nu} r^{-\nu}\\
\psi_-(r_*)&=e^{-i\omega r_*}\sum_{\nu=0}^{\infty} b^-_{\nu} r^{-\nu}
\end{align}
The coefficients are determined by the recurrence relations \cite{1997PThPh..98..359H},
\begin{align}
b_1^\pm
&=
\pm i \frac{l(l+1)}{2\omega} b_0^\pm \\
b_2^\pm
&=
\frac{1}{8\omega^2}
\Big[
-(l-1)l(l+1)(l+2)
\pm i 4\omega M
\Big] b_0^\pm \\
b_\nu^\pm
&=
\mp \frac{i}{2\omega \nu}
\Big[
\{(\nu-1)\nu - l(l+1)
\pm i 4\omega M(\nu-1)\} b_{\nu-1}^\pm \notag\\
&- 2M \{(\nu-1)(2\nu-3)-l(l+1)\} b_{\nu-2}^\pm \notag\\
&+ 4M^2(\nu-2)^2 b_{\nu-3}^\pm
\Big],
\quad (\nu \ge 3)
\end{align}

However, there are no incoming waves from infinity, so $C_+ = 0$, leaving only the $\psi_-$ term.\par

The boundary condition at the origin is $\psi = 0$. For numerical purposes, to cancel the singularity, we transform $\psi = r^{l+1} u(r)$ and expand
\begin{equation}
u = \sum_{\nu=0}^{\infty} a_\nu r^\nu
\label{A16}
\end{equation}
before performing the integration. Near the origin, the quantities can be expanded as
\begin{align}
T &= T_0 + T_2 r^2 + T_4 r^4 + \cdots , \tag{A4}\\
\Phi &= \Phi_0 + \Phi_2 r^2 + \Phi_4 r^4 + \cdots , \tag{A5}\\
\Psi &= \Psi_0 + \Psi_2 r^2 + \Psi_4 r^4 + \cdots . \tag{A6}
\end{align}
The coefficients $a_\nu$ are given as polynomials of these expansion coefficients \cite{1997PThPh..98..359H}. In the incompressible fluid model, the second- and fourth-order coefficients are already known. In general, the density and pressure of a dark star are not constant. However, near the center, the variations of $\epsilon$, $p$, and $\Phi$ are small (see figs.~\ref{profile2} and \ref{profile}). Therefore, in numerical calculations, for the first few points near the center, we simplify the computation by taking $T(=-\epsilon+3p)=T_0$ and $\Phi = \Phi_0$. $\Psi_4$ is trivial in principle. However, in this study, we focus only on the mode with $\ell = 0$ because we want to determine whether instability modes exist. Therefore, $\Psi_4$ can be neglected.
 This is because in the expressions for $a_\nu$, all terms containing $\Psi_4$ are multiplied by $\ell$, which vanish for $\ell = 0$.
The complex eigenfrequencies$ (\omega=\omega_R+i\omega_I)$ are obtained by searching for the zeros of the Wronskian,
\begin{equation}
W(u_{\rm in},u_{\rm out})=u_{\rm in}\frac{du_{\rm out}}{dr_{*}}-u_{\rm out}\frac{du_{\rm in}}{dr_{*}},
\end{equation}
in the complex-frequency plane. In our numerical calculation, the interior and exterior perturbation equations are integrated using a fourth-order Runge--Kutta method, and the complex eigenfrequencies are determined by locating the zeros of the Wronskian using the Müller method \cite{10.5555/573178}. We adopt the time dependence $e^{i\omega t}$, with the complex frequency written as $\omega=\omega_R+i\omega_I$. The perturbation therefore evolves as
\begin{equation}
e^{i\omega t}=e^{i\omega_R t}e^{-\omega_I t}.
\end{equation}
With this convention, modes with $\omega_I>0$ decay in time, whereas modes with $\omega_I<0$ grow exponentially and are therefore unstable. A tachyonic mode with $\omega^2<0$ has a purely imaginary frequency, because $-\frac{d^2}{dr_*^2} + V$ is Hermitian, which ensures that $\omega^2$ is real \cite{1997PThPh..98..359H}. Thus the growing branch corresponds to $\omega=-i|\omega_I|$. The real part $\omega_R$ is theoretically zero, but numerically it cannot be exactly zero; we take $\omega_R < O(10^{-10})$ as the criterion.

\begin{figure}
\centering
\includegraphics[width=\columnwidth]{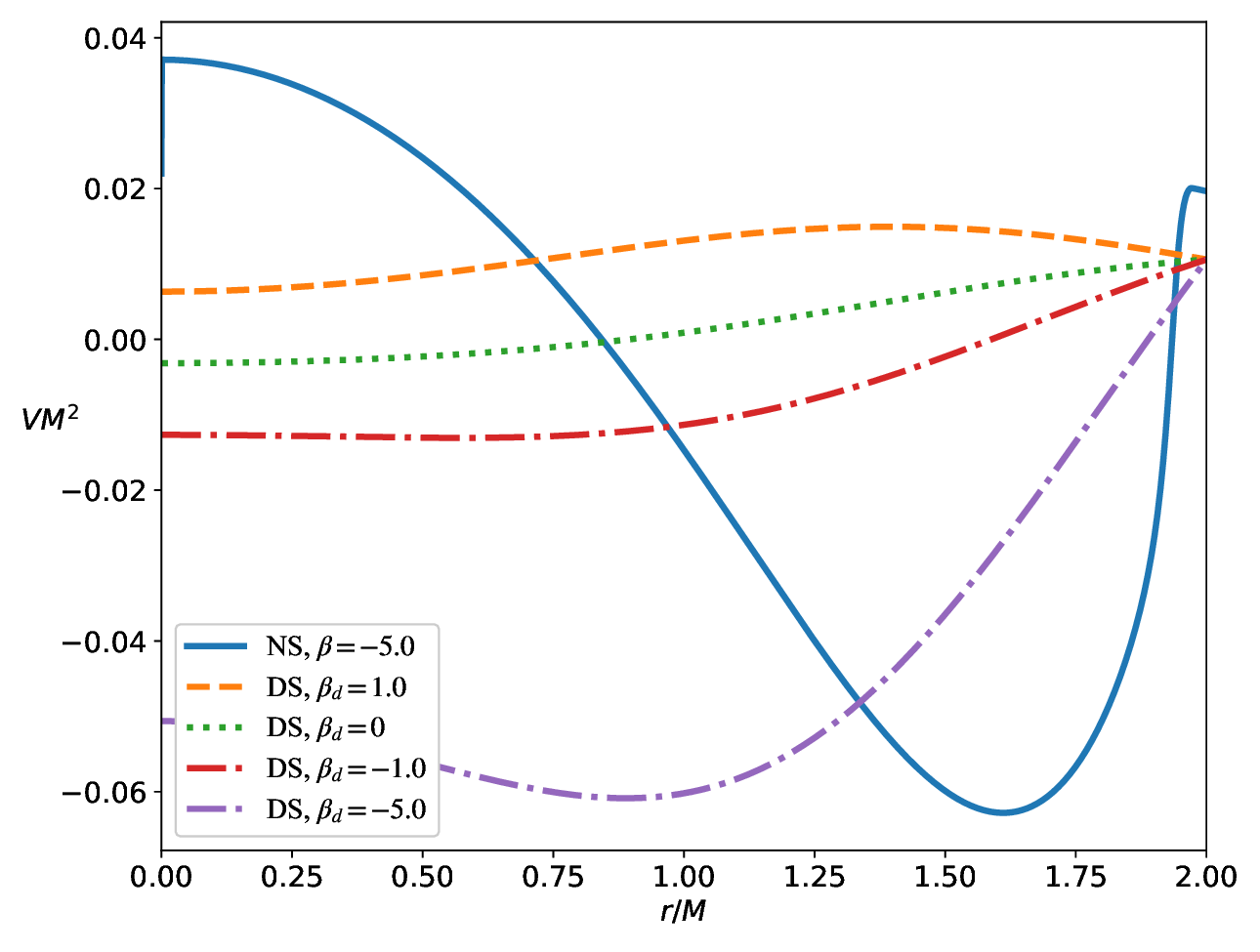}
\caption{\label{potenatial}Effective potential inside the dark star for $\beta_d = 1, 0, -1, -5$ and the neutron star for $\beta=-5$ with SLy EOS ($\ell=0$). The stellar mass is fixed at $M=2.0M_{\odot}$. The horizontal axis denotes the dimensionless radius $r/M$. }
\end{figure}

\section{\label{NR}Tachyonic instability in dark stars}

\subsection{Pure dark matter stars}

For the system to have unstable modes, bound states must exist, which requires the effective potential to have negative regions. Fig.~\ref{potenatial} shows the effective potential for dark stars and a neutron star. When $\beta_d = 1$, the effective potential is positive throughout the star ($r/M \le 2$), whereas for $\beta_d \le 0$, negative regions appear, and the potential well becomes deeper as $|\beta_d|$ increases. Furthermore, compared to neutron stars, the potential well is wider and deeper even at the star's core. Unstable modes exist only when $\beta_d$ takes negative values.\par

\begin{figure}
\centering
\includegraphics[width=\columnwidth]{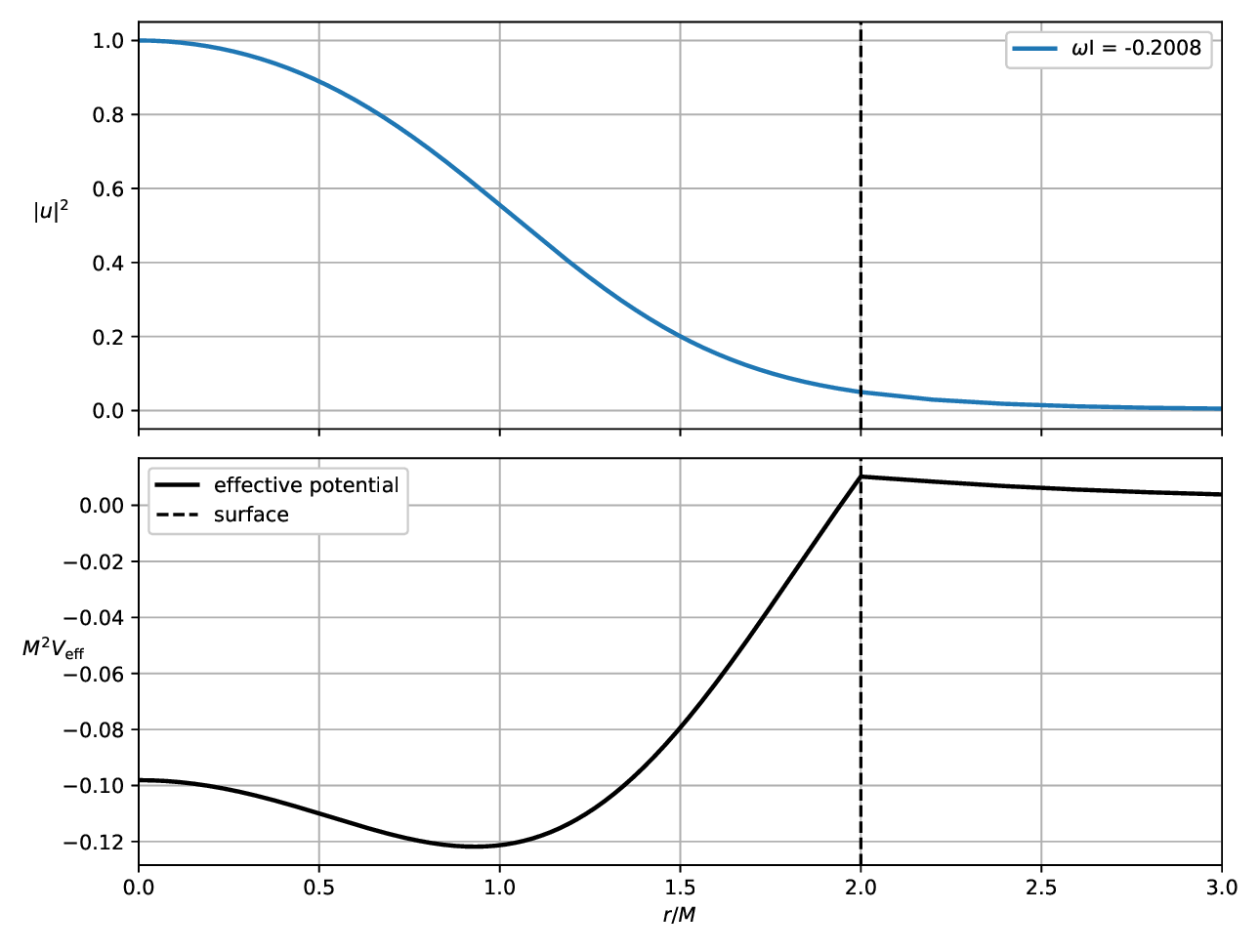}
\caption{Upper panel: The distribution of $|u|^2$ corresponding to the mode with $\omega_I=-0.2008$ at $\beta_d=-10$. The distribution is normalized so that the maximum value is $|u|^2=1$. Lower panel: Effective potential at $\beta_d=-10$. The black vertical dashed line indicates the stellar surface. Stellar mass is $M = 2.0 M_{\odot}$. Dark matter particle parameters are $(\mu,\kappa)=(4.1,9.2)$.}
\label{M2m4.1k9.2b-10psi}
\end{figure}

\begin{figure}
\centering
\includegraphics[width=\columnwidth]{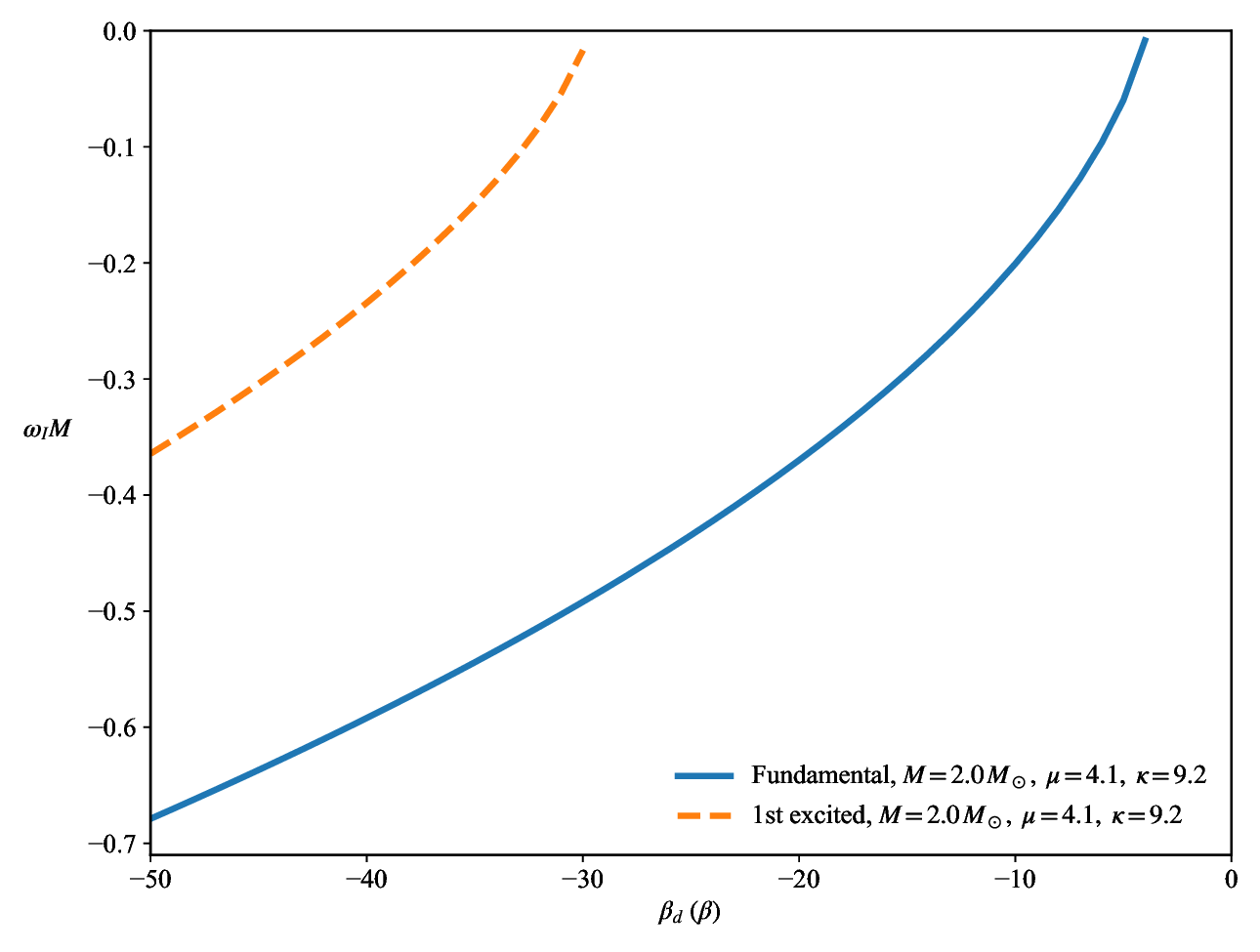}
\caption{\label{M2m4.1k9.2omega}Variation of $\omega_I$ for a dark star. Stellar mass is $M = 2.0 M_{\odot}$. Dark matter particle parameters are $(\mu,\kappa)=(4.1, 9.2)$.}
\end{figure}

\begin{figure}
\centering
\includegraphics[width=\columnwidth]{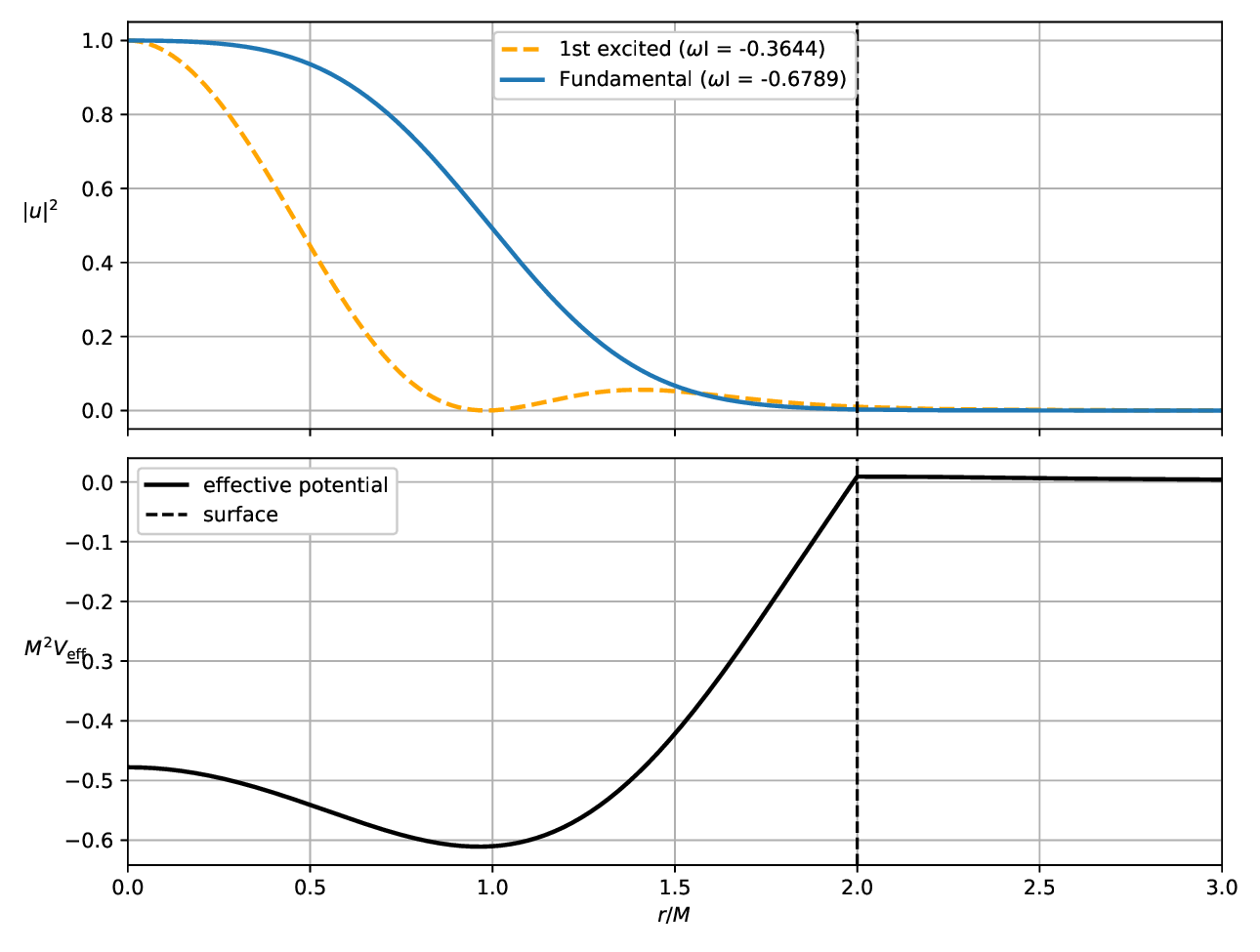}
\caption{\label{M2m4.1k9.2b-50psi}Upper panel: The distribution of $|u|^2$ corresponding to the fundamental mode ($\omega_I=-0.6789$) and the first excited mode ($\omega_I=-0.3644$) at $\beta_d=-50$. Each mode is normalized independently so that its maximum value is $|u|^2=1$. Therefore, the amplitudes of $|u|^2$ cannot be compared between different modes. Lower panel: Effective potential at $\beta_d=-50$. The black vertical dashed line indicates the stellar surface. Stellar mass is $M = 2.0 M_{\odot}$. Dark matter particle parameters are $(\mu,\kappa)=(4.1,9.2)$.}
\end{figure}

Fig.~\ref{M2m4.1k9.2b-10psi} shows the distribution of $|u|^2$ at $\beta_d=-10$ for a pure dark star with $(\mu,\kappa)=(4.1,9.2)$. As can be seen from the figure, a well in the effective potential has been created, and the scalar distribution is localized around the star’s center, with part of it extending beyond the star's boundaries.
Fig.~\ref{M2m4.1k9.2omega} shows the relationship between $\beta_d$ and $\omega_I$ for a pure dark star with $(\mu,\kappa)=(4.1,9.2)$, and mass $M=2.0M_{\odot}$. For relatively large values of $\beta_d$, the dark star possesses only one unstable mode, whereas for smaller values of $\beta_d$ it exhibits two unstable modes. Fig.~\ref{M2m4.1k9.2b-50psi} shows the corresponding distribution at $\beta_d=-50$. At $\beta_d=-10$, only the fundamental mode exists, whereas at $\beta_d=-50$ an additional first excited mode appears. This excited mode corresponds to the mode that emerges as $\beta_d$ becomes more negative and the effective potential becomes deeper. 

The growth time is $\tau=1/|\omega_I|=M/(|\omega_I|M)$. For example, when $\beta = -4$, in  the case of $M=2.0M_{\odot}$, $\tau \simeq 10^{-3}s$. This indicates that tachyonic instability immediately occurs in the region with $\beta_d < -4$ when the dark matter star is at the typical mass scale of neutron stars ($\simeq 2.0M_{\odot}$).\par

\begin{figure}
\centering
\includegraphics[width=\columnwidth]{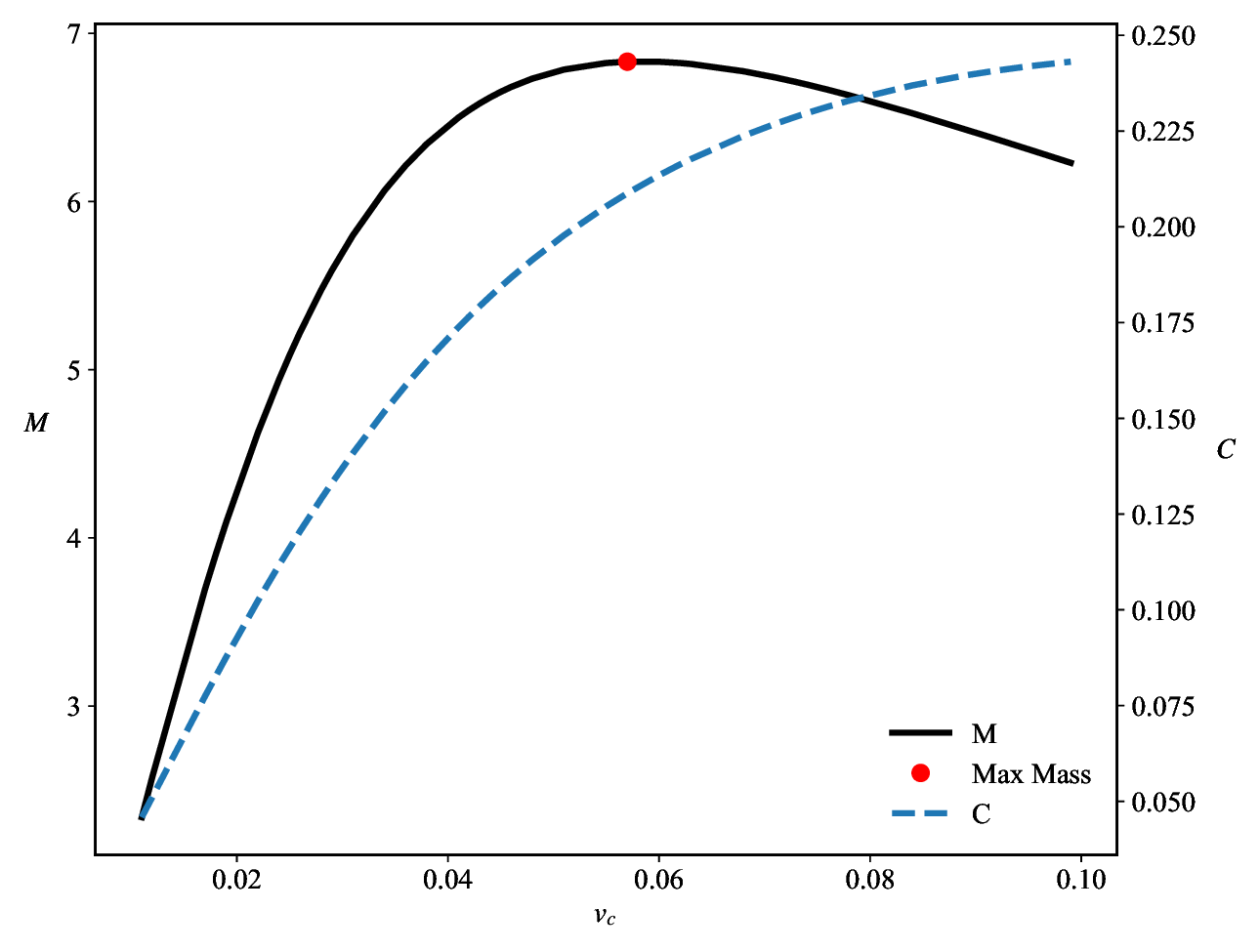}[th]
\caption{\label{1-1mc}Relationship between $v_c$ and the stellar mass and compactness for dark stars with $(\mu,\kappa)=(1.0,1.0)$. The left vertical axis denotes the stellar mass, while the right vertical axis denotes the compactness. The red circle indicates the maximum-mass configuration.}
\end{figure}

The next question is which dark stars exhibit the strongest instability. In general, studies of spontaneous scalarization suggest that higher compactness tends to lead to greater instability. Therefore, we investigated the relationship between compactness and the tachyonic instability. Fig.~\ref{1-1mc} shows the mass and compactness of dark stars with $(\mu,\kappa)=(1.0,1.0)$. The horizontal axis is the parameter $v_c$ (Eq.~\ref{eq:vc}), which is related to the Fermi momentum at the stellar center. As $v_c$ increases, both the stellar mass and compactness increase. However, the compactness reaches its maximum only after the stellar mass has already attained its maximum value. As is well known, stars become dynamically unstable in the region where $dM/dv_c<0$. Therefore, the dark stars to the right of the maximum-mass point in the figure are unstable. Consequently, the maximum compactness is not realized by stable dark stars.

\begin{figure}
\centering
\includegraphics[width=\columnwidth]{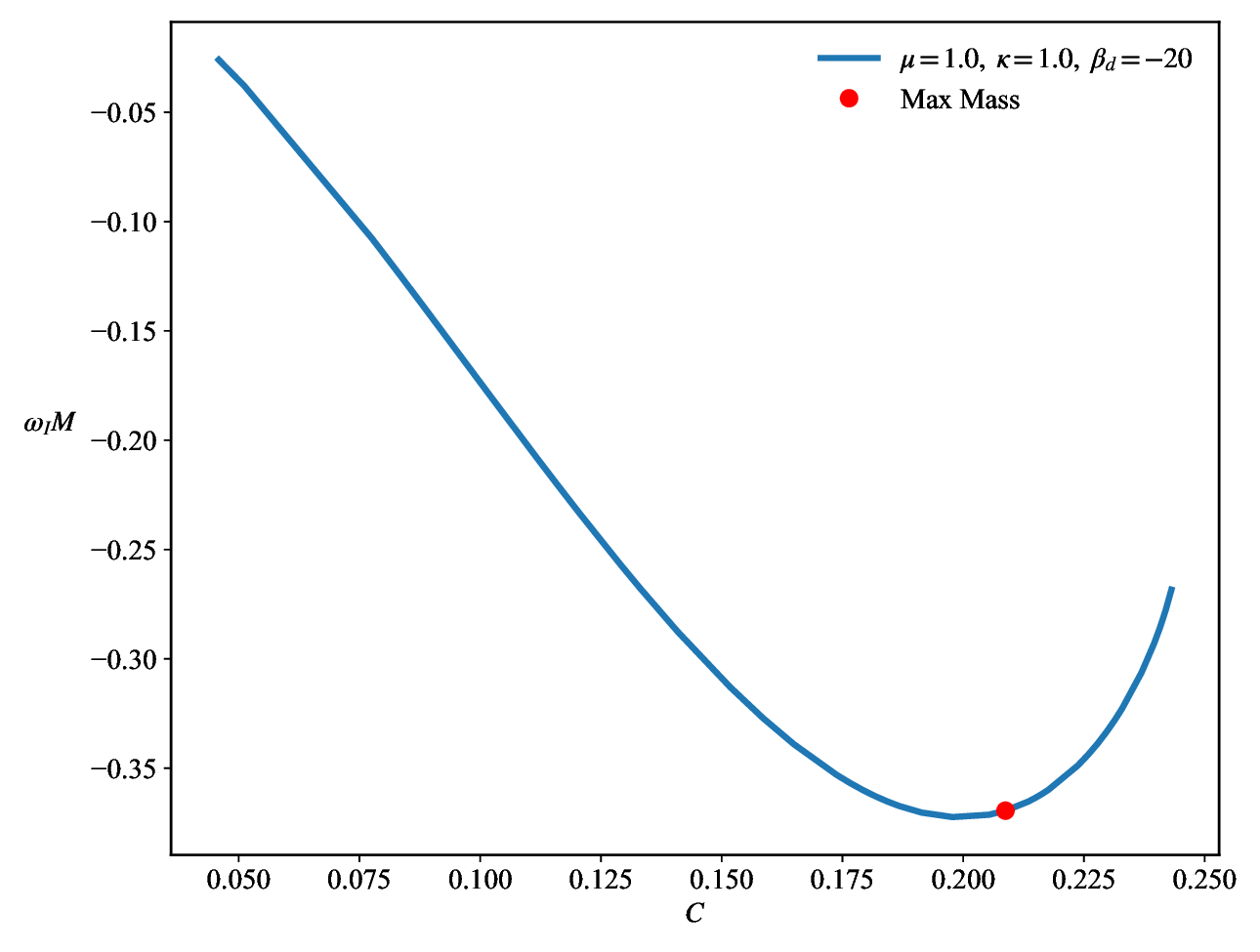}
\caption{\label{1-1b-20comega}Relationship between $\omega_I$ of the unstable mode at $\beta_d=-20$ and $v_c$ for dark stars with $(\mu,\kappa)=(1.0,1.0)$. The red circle indicates the maximum-mass configuration.}
\end{figure}

\begin{figure}
\centering
\includegraphics[width=\columnwidth]{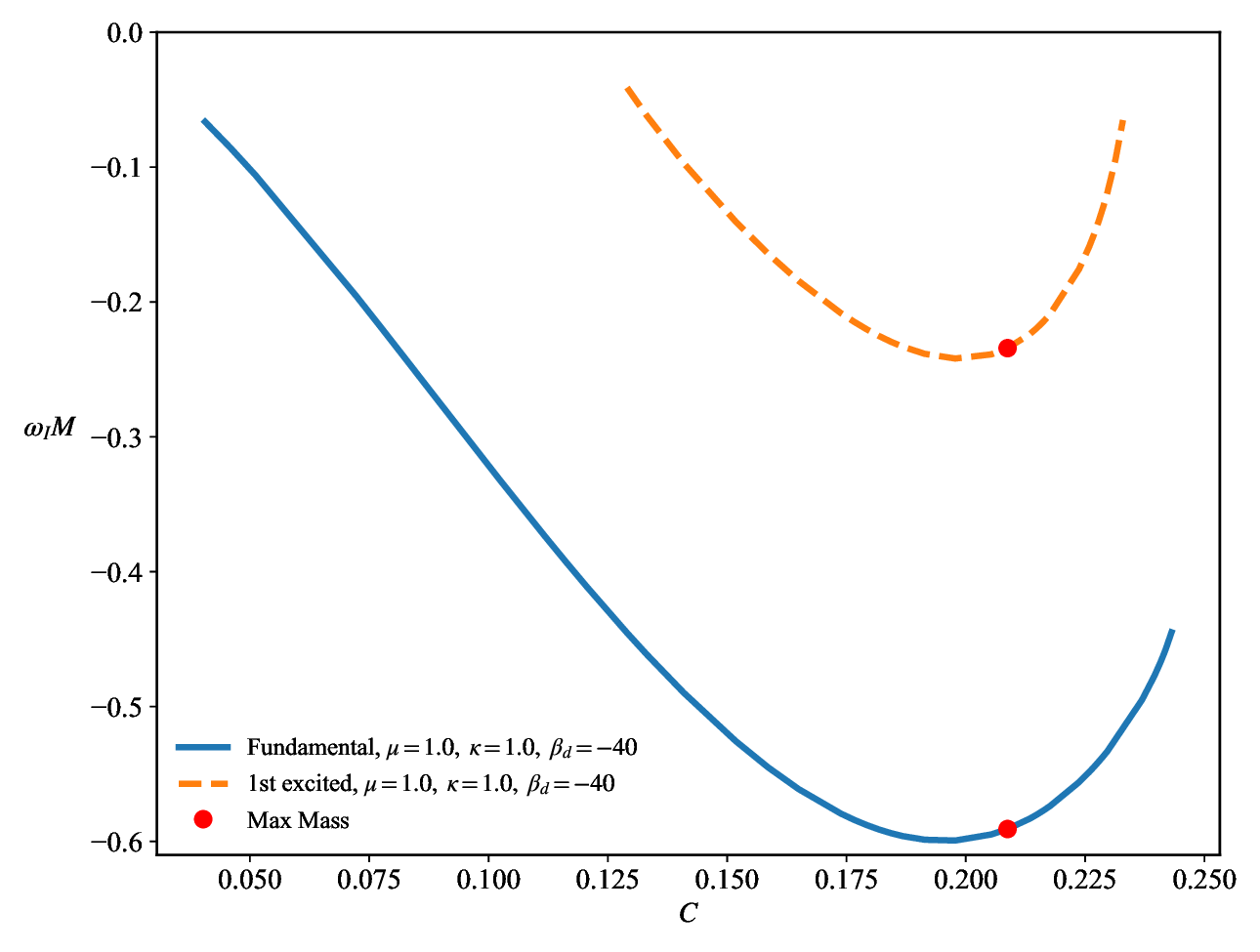}
\caption{\label{1-1b-40comega}Relationship between $\omega_I$ of the unstable modes at $\beta_d=-40$ and $v_c$ for dark stars with $(\mu,\kappa)=(1.0,1.0)$. In addition to the fundamental mode, the first excited mode is also present. The red circle indicates the maximum-mass configuration.}
\end{figure}

Fig.~\ref{1-1b-20comega} and \ref{1-1b-40comega} show the relationship between the compactness and $\omega_I$ for $\beta_d=-20$ and $\beta_d=-40$, respectively. In both figures, the upper-left corner corresponds to low values of $v_c$, and $v_c$ increases toward the lower-right corner. As $v_c$ increases, $|\omega_I|$ first increases, reaches a maximum, and then decreases. The red circles indicates the maximum-mass configurations. Configurations beyond these points are dynamically unstable. Therefore, the strongest scalar instability occurs not for the most compact star, but rather for a stable star whose mass is nearly the maximum mass along the equilibrium sequence for a given set of $(\mu,\kappa)$.

It should be noted that the sound speed at the stellar center,
\begin{equation}
c_s^2=\frac{dp}{d\epsilon},
\end{equation}
satisfies the causality condition $c_s^2<1$ over the entire range of $v_c$ shown in the figures.

\begin{figure}
\centering
\includegraphics[width=\columnwidth]{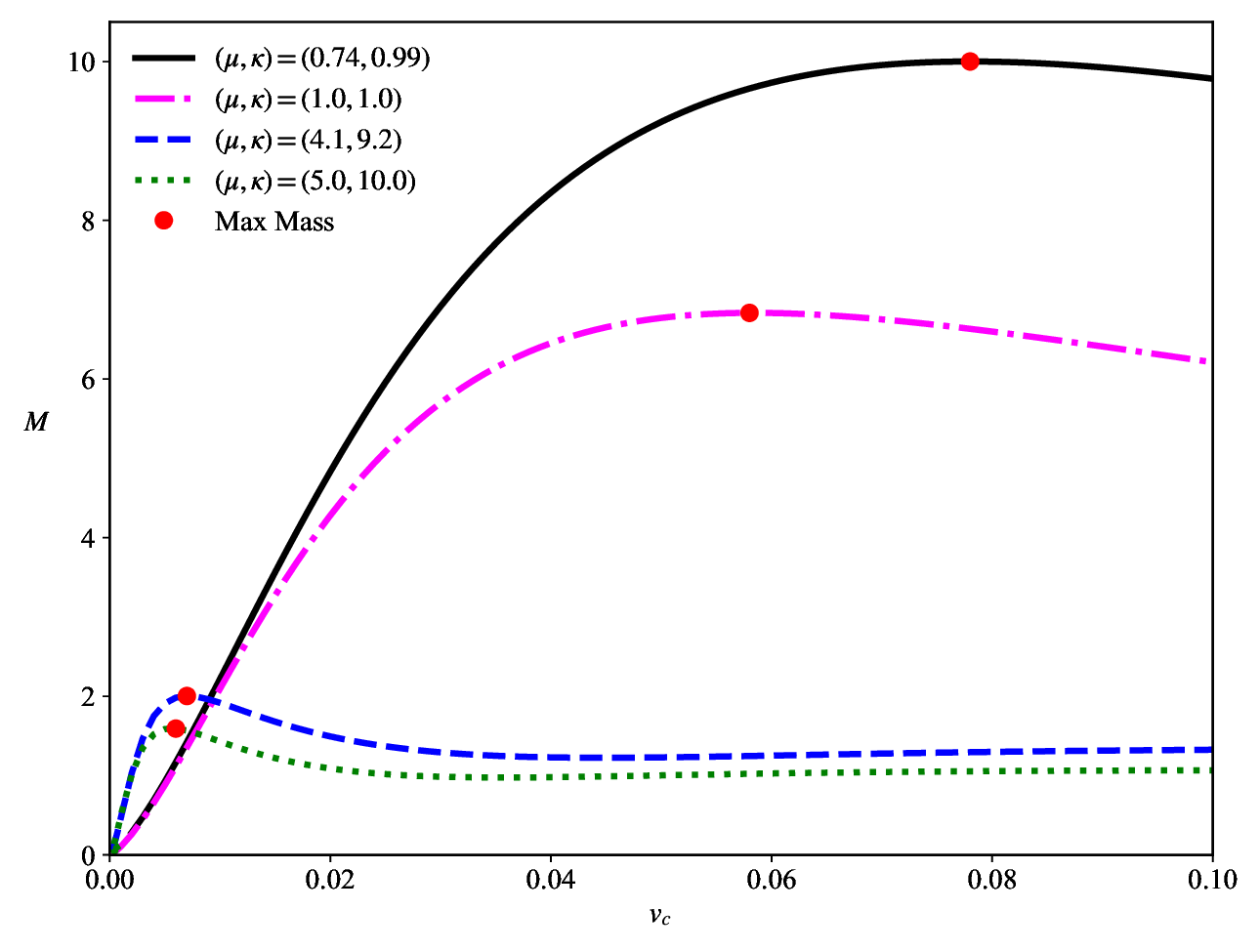}
\caption{\label{mc} The stellar sequence of $v_c$ for each $(\mu,\kappa)=(0.74,0.99),(1.0,1.0),(4.1,9.2),(5.0,10.0)$.The red circle indicates the maximum-mass configulation for eqch $(\mu,\kappa)$ sequence}
\end{figure}

\begin{figure}
\centering
\includegraphics[width=\columnwidth]{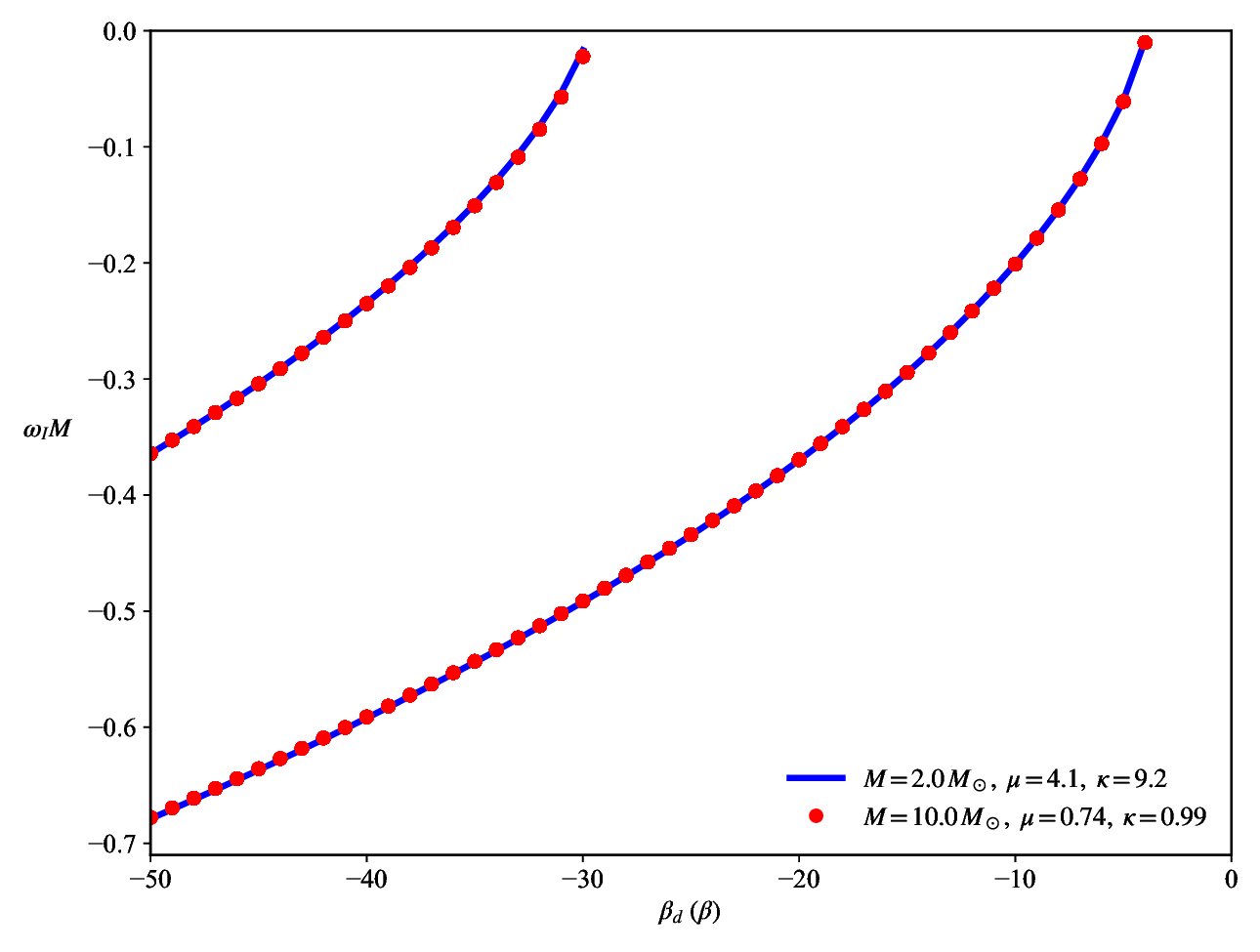}
\caption{\label{com}Comparison of $\omega_I$ between the dark star with $M=2.0M_{\odot}$, $(\mu,\kappa)=(4.1,9.2)$, and the dark star with $M=10.0M_{\odot}$, $(\mu,\kappa)=(0.74,0.99)$. The former is shown by the blue curve, and the latter by the red circles.}
\end{figure}

\begin{figure}
\centering
\includegraphics[width=\columnwidth]{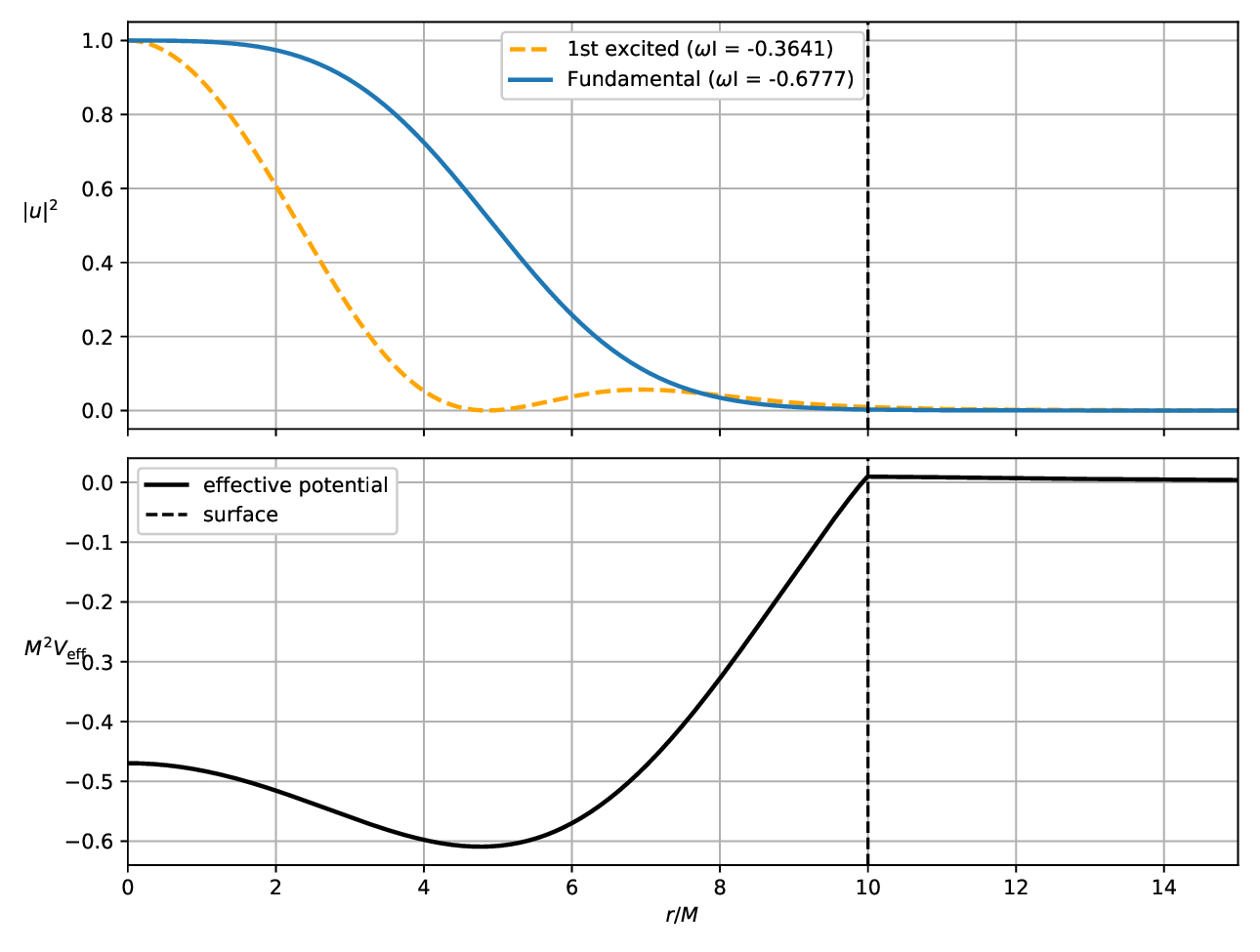}
\caption{\label{M10m074k099b-50psi}Upper panel: The distribution of $|u|^2$ corresponding to the fundamental mode ($\omega_I=-0.6777$) and the first excited mode ($\omega_I=-0.3641$) at $\beta_d=-50$. Each mode is normalized independently so that its maximum value is $|u|^2=1$. Therefore, the amplitudes of $|u|^2$ cannot be compared between different modes. Lower panel: Effective potential at $\beta_d=-50$. The black vertical dashed line indicates the stellar surface. Stellar mass is $M = 10.0 M_{\odot}$. Dark matter particle parameters are $(\mu,\kappa)=(0.74,0.99)$.}
\end{figure}

\begin{figure}
\centering
\includegraphics[width=\columnwidth]{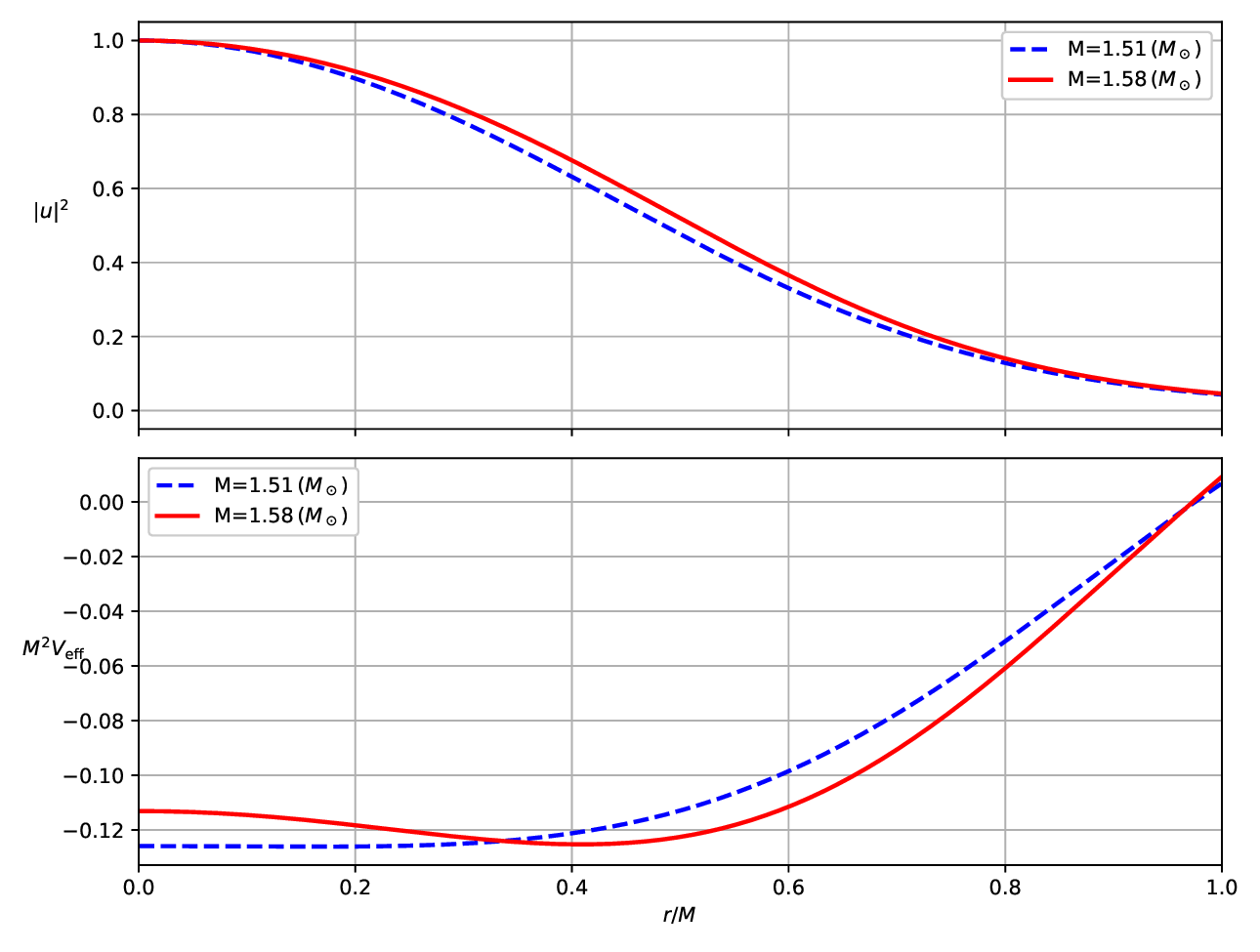}
\caption{\label{psicom1}Upper panel: Comparison of the $|u|^2$ distributions for stars with $M=1.51$, $C=0.1753$, and $M=1.58$, $C=0.1993$, corresponding to the mode at $\beta_d=-10$. The distributions are normalized so that the maximum value is $|u|^2=1$. Unlike the other figures, the stellar surfaces of the two stars are aligned for comparison. Lower panel: Effective potentials of the two stars at $\beta_d=-10$.}
\end{figure}

\begin{figure}
\centering
\includegraphics[width=\columnwidth]{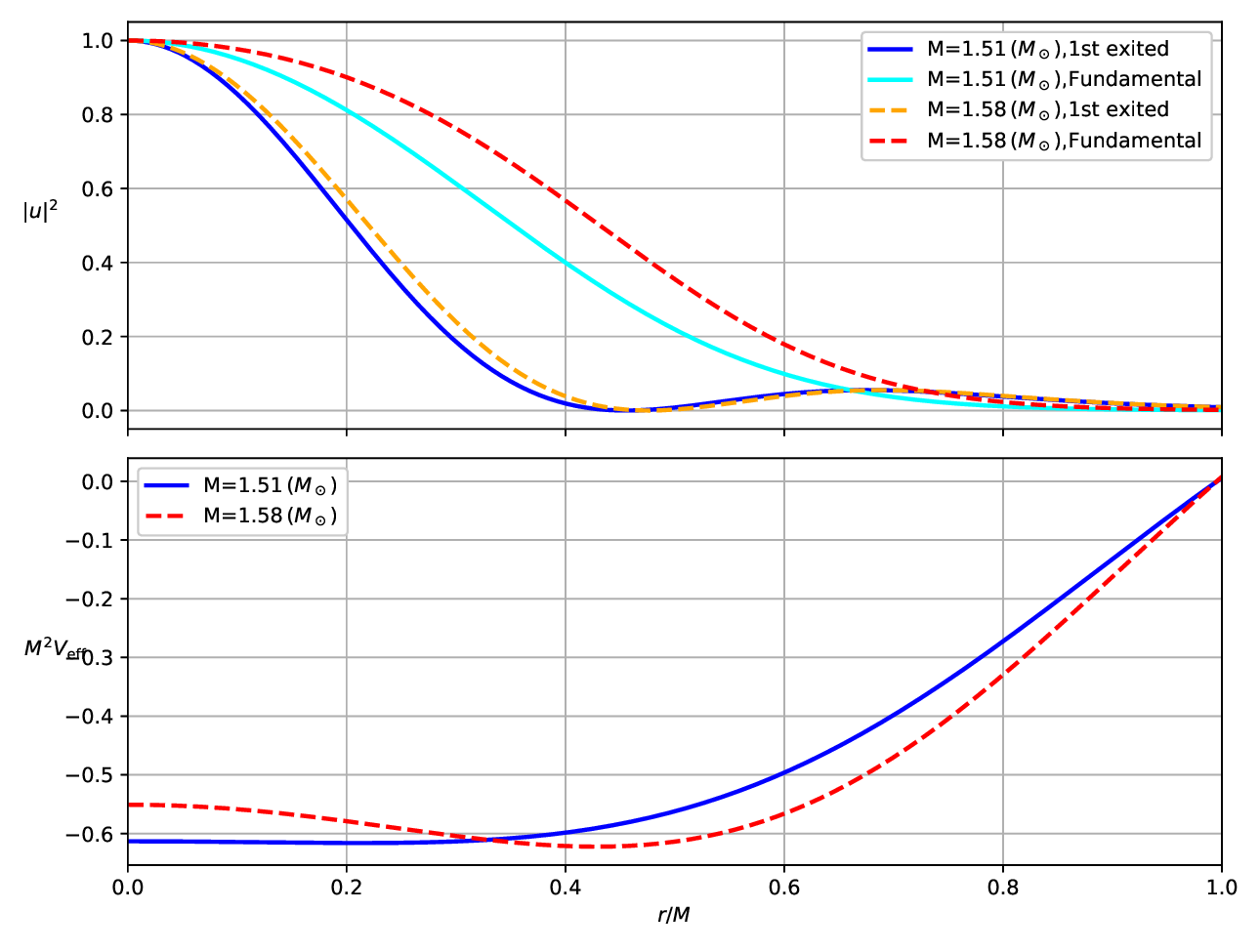}
\caption{\label{psicom2}Upper panel: Comparison of the $|u|^2$ distributions for stars with $M=1.51$, $C=0.1753$, and $M=1.58$, $C=0.1993$, corresponding to the fundamental mode and the first excited mode at $\beta_d=-50$. Each mode is normalized independently so that its maximum value is $|u|^2=1$. Unlike the other figures, the stellar surfaces of the two stars are aligned for comparison. Lower panel: Effective potentials of the two stars at $\beta_d=-50$.}
\end{figure}

Fig.~\ref{mc} shows the stellar mass sequences for each  $(\mu,\kappa)$ stars as $v_c$ varies. The red circles indicates the points of maximum mass. In fact, the star of $(\mu,\kappa)=(4.1,9.2),M=2.0M_{\odot}$ is nearly at its maximum mass on the mass sequence.

Fig.~\ref{com} compares two dark stars: one with $(\mu,\kappa)=(4.1,9.2)$, and $M=2.0M_{\odot}$, and another with $(\mu,\kappa)=(0.74,0.99)$, and $M=10.0M_{\odot}$. These stars are both near the maximum mass point of their respective sequences (Fig.~\ref{mc}). Consequently, their unstable mode frequencies are remarkably similar. Furthermore, despite their significantly different masses and radii, the scalar-field distributions $|u|^2$ and the shapes of the effective potentials are also very similar (Figs.~\ref{M2m4.1k9.2b-50psi} and \ref{M10m074k099b-50psi}). These results indicate that, regardless of the $(\mu, \kappa)$ combination, the maximum growth rate is nearly the same and is attained near the maximum-mass configuration. Furthermore, for all $(\mu, \kappa)$ pairs, the compactness near the maximum mass was $C \sim 0.21$. Since this is the endpoint of the stable branch for dark stars, the maximum compactness of a stable star is $C \sim 0.21$.\par

Although the star's mass changes, as it transitions to its maximum mass, the effective potential and the scalar distribution change accordingly. Figs.~\ref{psicom1} and \ref{psicom2} compare two stars with $(\mu,\kappa)=(5.0,10.0)$ but different stellar mass, $M=1.51M_{\odot}$ and $1.58M_{\odot}$, respectively. $1.58M_{\odot}$ is near the maximum mass point in this sequence. Even though the two stars are composed of the same dark matter particles and differ only slightly in mass, the effective potential is highly sensitive to change of mass. As the minimum of the effective potential shifts away from the stellar center, the distribution of $|u|^2$ changes accordingly and follows in that direction. \par

\begin{figure}
\centering
\includegraphics[width=\columnwidth]{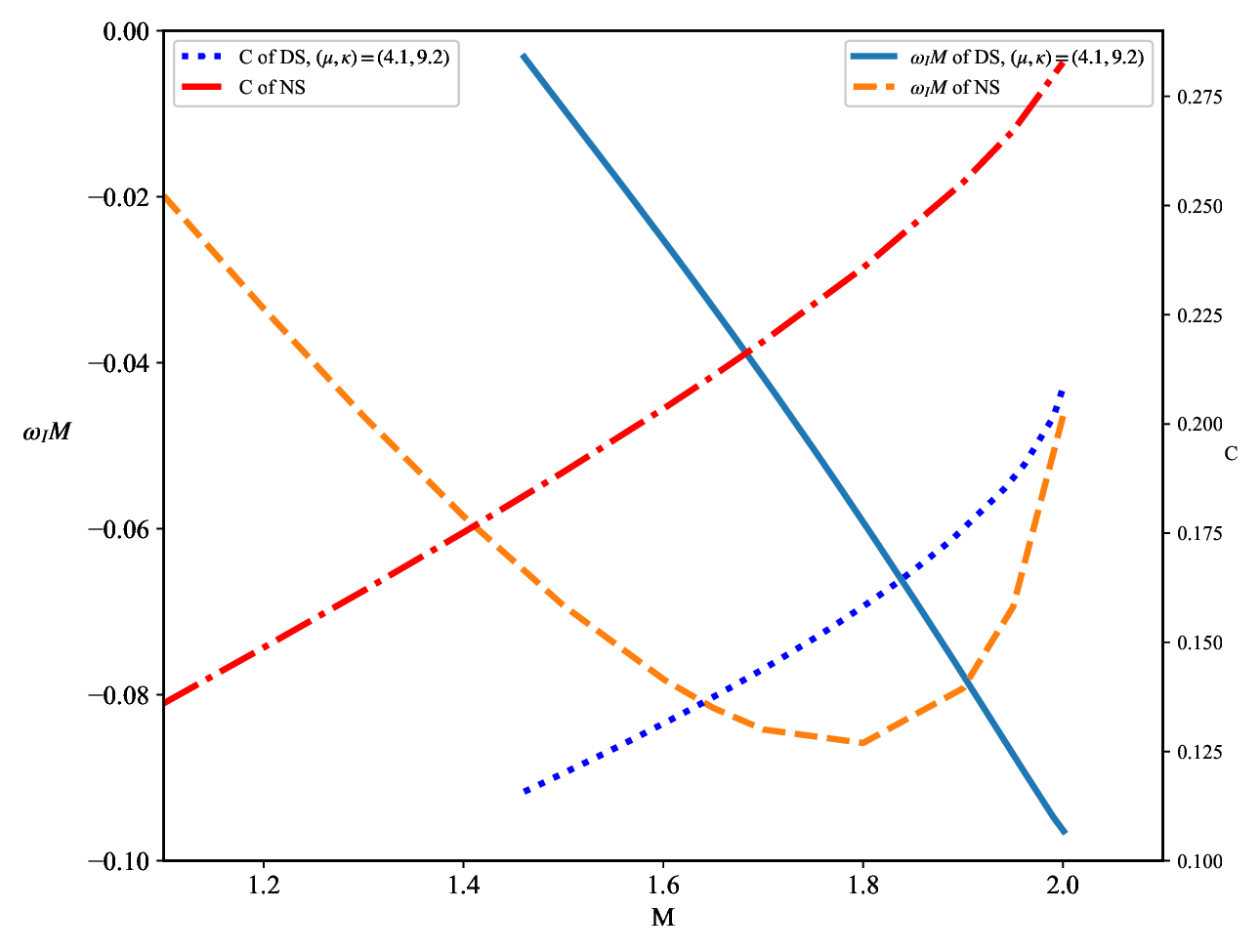}
\caption{\label{dsns}Comparison of $\omega_I$ and $C$ for dark stars and neutron stars. The horizontal axis represents stellar mass. Dark matter particle parameters are $(\mu,\kappa)=(4.1,9.2)$. For neutron stars with the SLy EOS. $\beta_d=\beta=-6.0$.}
\end{figure}

\begin{figure}
\centering
\includegraphics[width=\columnwidth]{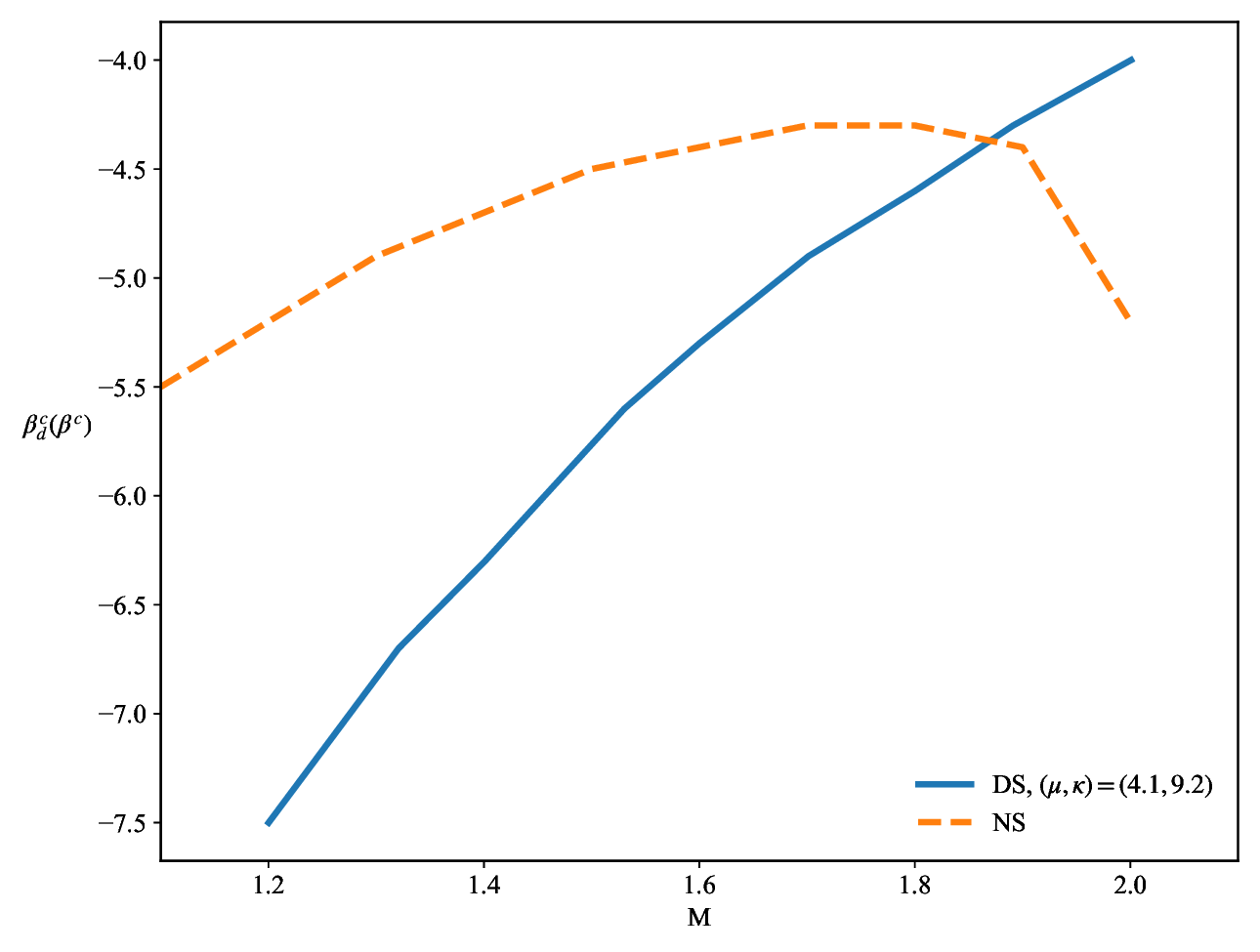}
\caption{\label{bb}Comparison of $\beta_d^c$ and $\beta^c$. The horizontal axis represents stellar mass.  Dark matter particle parameters are $(\mu,\kappa)=(4.1,9.2)$. For neutron stars with the SLy EOS. }
\end{figure}

Fig.~\ref{dsns} shows the imaginary part $\omega$ and compactness $C$ for dark stars and neutron stars. For dark stars, the figure shows the sequence of $(\mu,\kappa)=(4.1,9.2)$. For neutron stars, we adopt the SLy equation of state (EOS) \cite{2001A&A...380..151D,Haensel2001,2003EAS.....7..249H}.
Fig.~\ref{dsns} shows that, in a dark star, both $|\omega_I|$ and $C$ increase as the star's mass increases; however, in a neutron star, while compactness increases, $|\omega_I|$ reaches a maximum and then decreases. Consequently, near the point of maximum mass $(\sim 2.0M_{\odot})$, the $|\omega_I|$ values are reversed in the dark star. In other words, while neutron stars are generally more compact than dark stars, that alone did not result in a larger $|\omega_I|$.
On the other hand, if we denote the maximum value of $\beta_d$ (or $\beta$) at which the tachyonic instability occurs as $\beta_d^c$ (or $\beta^c$), Fig.~\ref{bb} shows $\beta_d^c$ ($\beta^c$) for various masses of dark stars and neutron stars. Here, too, for low-mass stars, $\beta^c$ is greater than $\beta_d^c$ for neutron stars, but this relationship reverses as the star approaches its maximum mass. As these figures show, it is difficult to provide a unified explanation for tachyonic instabilities in dark stars and neutron stars based solely on compactness.
\par

\begin{figure}
\centering
\includegraphics[width=\columnwidth]{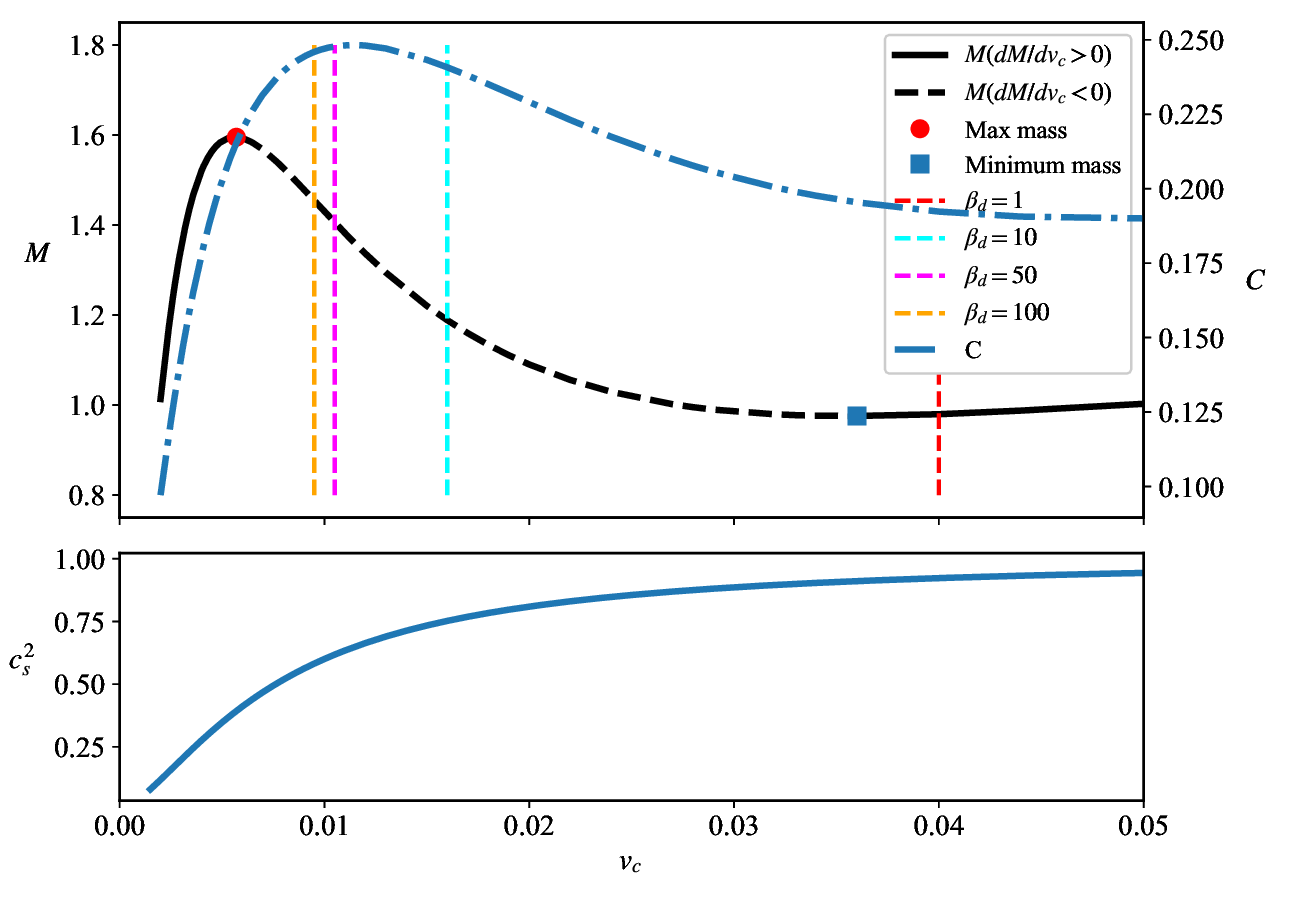}
\caption{\label{5-10mccs}Upper panel: Relationship between $v_c$ and the stellar mass and compactness for dark stars with $\mu=5.0$ and $\kappa=10.0$. The left vertical axis denotes the stellar mass, while the right vertical axis denotes the compactness. The red circle and the blue square indicate the maximum- and minimum-mass configurations, respectively. The black curve represents the stellar mass. The dashed segment between the red circle and the blue square corresponds to the region with $dM/dv_c<0$, while the remaining solid segments correspond to $dM/dv_c>0$. Each vertical dashed line indicates the minimum value of $v_c$ for which an unstable mode exists at the corresponding value of $\beta_d$. For example, unstable modes with $\beta_d=100$ exist only to the right of the line labeled $\beta_d=100$. Lower panel: Central sound speed as a function of $v_c$.}
\end{figure}

\begin{figure}
\centering
\includegraphics[width=\columnwidth]{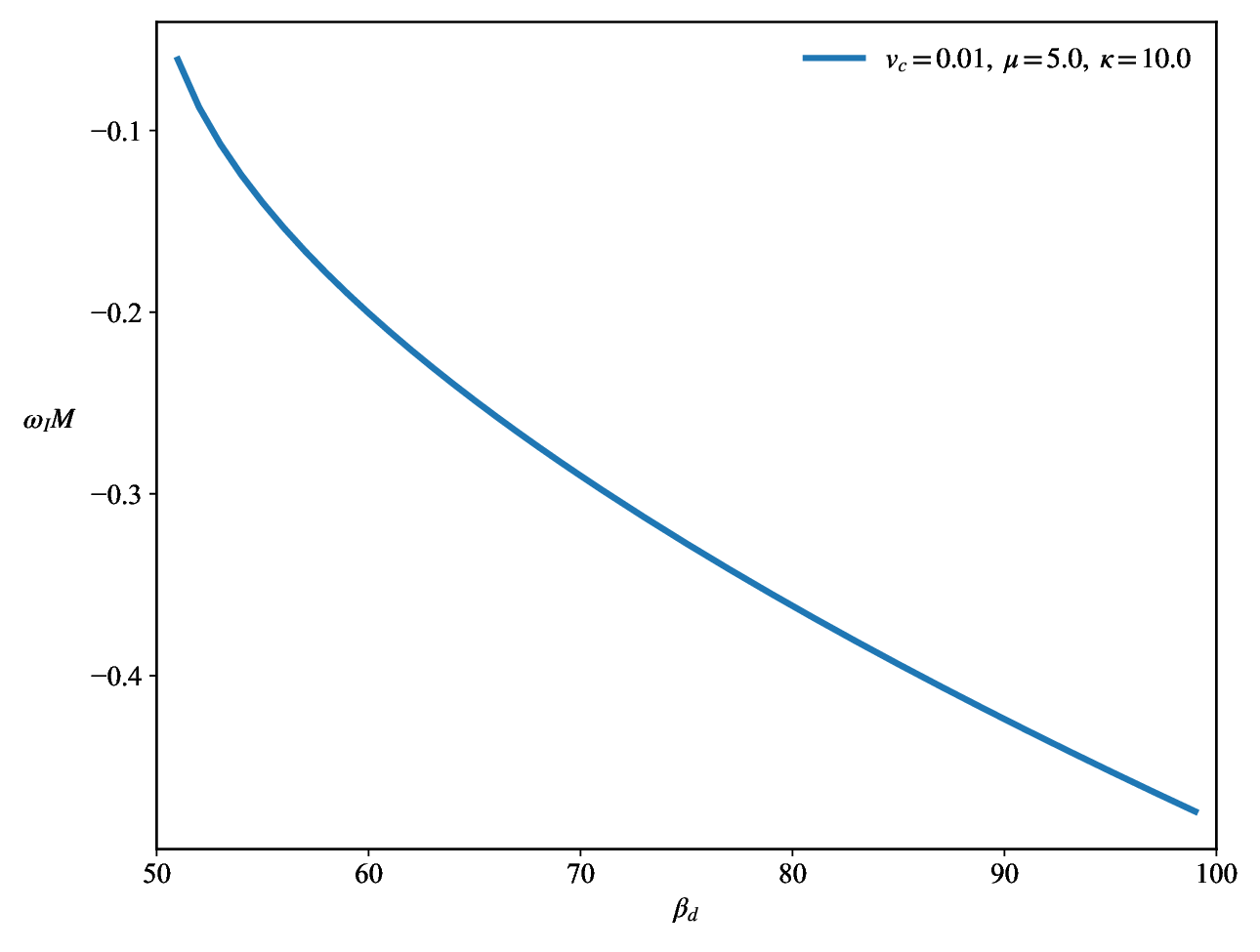}
\caption{\label{vc0.01b+}Imaginary part of the unstable mode, $\omega_I$, as a function of $\beta_d$ for a dark star with $(\mu,\kappa)=(5.0,10.0)$, and $v_c=0.01$.}
\end{figure}

\begin{figure}
\centering
\includegraphics[width=\columnwidth]{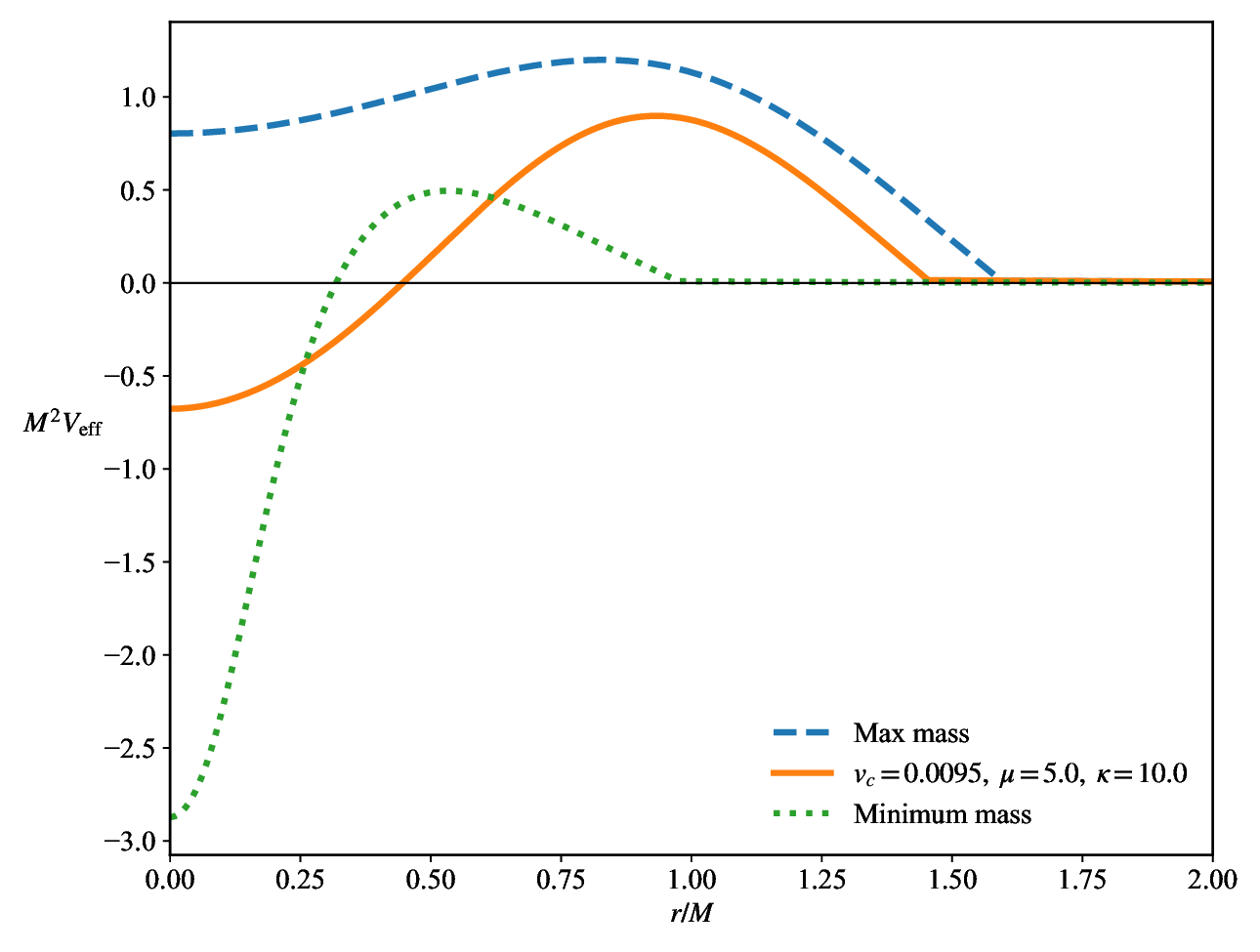}
\caption{\label{b+100potential}Effective potential for $\beta_d=100$ for the maximum-mass configuration, the minimum-mass configuration, and the configuration with $v_c=0.0095$, for dark stars with $(\mu,\kappa)=(5.0,10.0)$.}
\end{figure}

So far, we have focused on the region $\beta_d < 0$. For stars composed of ordinary matter, such as neutron stars, tachyonic instability occurs only for certain negative values of $\beta$, because the effective mass is proportional to the product of $\beta$ and the trace of the energy–momentum tensor $T$. Except for some EOSs with high central density, $T$ is generally negative. While $\beta_d$ is not necessarily equal to $\beta$, Fig. \ref{M2m4.1k9.2omega}  indicates that the system does not possess unstable modes for $\beta_d > 0$. This is because a stable dark star, like a neutron star, has a structure with $T < 0$. In our self-interacting dark matter model, the structure of the star is primarily determined by the parameter $v_c$ related to the central Fermi momentum of dark matter particles at the center. If $v_c$ increases, the central density can become high enough that $T > 0$. 
Fig.~\ref{5-10mccs} shows the stellar mass and compactness as functions of $v_c$. As $v_c$ increases, the stellar mass first reaches a maximum and then decreases. Since $dM/dv_c<0$ along this branch, it is dynamically unstable. After the mass reaches its minimum, however, $dM/dv_c>0$ again. Therefore, based solely on the mass criterion, this branch cannot be regarded as dynamically unstable. In the mass curve, the dashed segment indicates the unstable branch. Nevertheless, in part of this region and throughout the region beyond it, the trace satisfies $T>0$, allowing the effective potential to develop a potential well even for $\beta_d>0$ (Fig.~\ref{b+100potential}). Consequently, unstable modes can exist if $\beta_d$ is sufficiently large (Fig.~\ref{vc0.01b+}). The vertical dashed lines in Fig.~\ref{5-10mccs} indicates the minimum value of $v_c$ at which an unstable mode appears for $\beta_d=1$, 10, 50, and 100. Therefore, unstable modes exist to the right of each corresponding dashed line.

As discussed above, since $dM/dv_c>0$ beyond the minimum-mass configuration, this branch is not identified as dynamically unstable by the mass criterion, which diagnoses radial stability only. However, this interpretation requires caution. The lower panel of Fig.~\ref{5-10mccs} shows the sound speed, $c_s^2={dp}/{d\epsilon}$ as a function of $v_c$. The causality condition, $c_s^2<1$, is satisfied over the entire parameter range shown. However, after passing the minimum-mass configuration, the sound speed reaches $c_s^2\sim0.9$ for $v_c\gtrsim0.04$, where unstable modes already exist for $\beta_d=1$. Such configurations therefore correspond to extremely dense stars with very stiff equations of state. This stiffness originates from the repulsive Yukawa self-interaction of the dark matter. Because self-interacting dark matter possesses an intrinsic mass scale and Yukawa interaction, its equation of state is not immediately constrained by the conformal-limit value, $c_s^2=1/3$. Nevertheless, for small positive values of $\beta_d$, unstable modes appear only in these extremely stiff stars. Even for larger positive values of $\beta_d$, unstable modes occur only in dynamically unstable stellar configurations.

\begin{figure}
\centering
\includegraphics[width=\columnwidth]{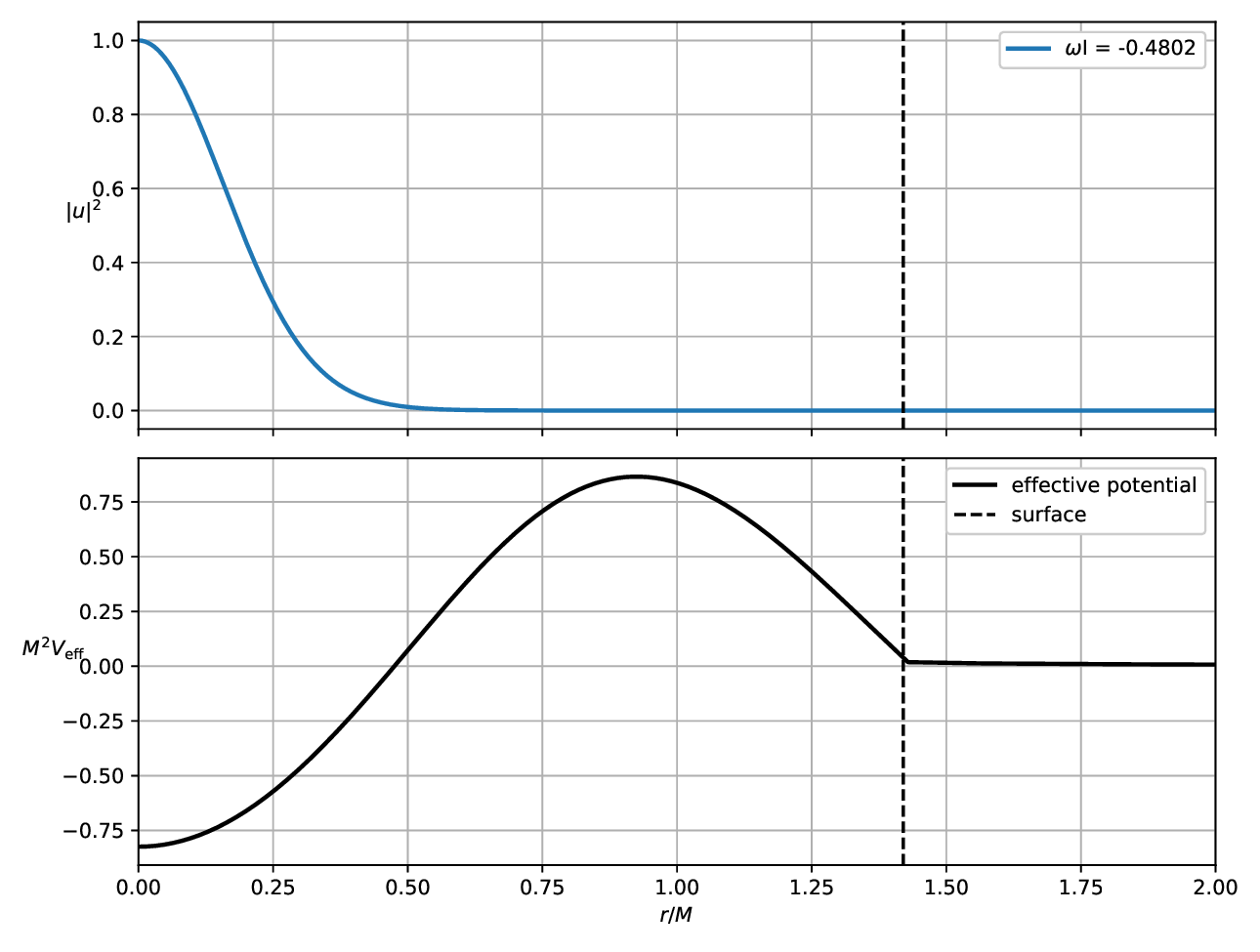}
\caption{\label{5-10b+100vc0.01psi}Upper panel: The distribution of $|u|^2$ corresponding to the unstable mode at $\beta_d=100$ for a dark star with $(\mu,\kappa)=(5.0,10.0)$, and $v_c=0.01$. Lower panel: Corresponding effective potential. The black vertical dashed line indicates the stellar surface.}
\end{figure}

Furthermore, the nature of these unstable modes differs from those found in dynamically stable stars. Fig.~\ref{5-10b+100vc0.01psi} shows the distribution of $|u|^2$ and the corresponding effective potential for $v_c=0.01$ and $\beta_d=100$. Compared with the stable configurations discussed previously, the effective potential does not develop a broad well inside the star and is significantly shallower. As a result, the scalar-field distribution is strongly localized near the stellar center and does not extend toward the stellar surface, in contrast to the behavior found in dynamically stable stars.

\subsection{doubly-degenerate stars}
In this section, we consider mixed stars composed of dark matter and baryonic matter. Such objects include both configurations in which baryonic gas accumulates at the center of a dark star and those in which dark matter is captured inside a neutron star. Collectively, these objects are referred to as doubly-degenerate stars (DDSs). The former consists of a mixed dark matter--baryon core surrounded by a dark matter envelope, whereas the latter consists of a mixed core surrounded by a baryonic envelope. As in the case of pure dark stars, we assume a static, spherically symmetric spacetime described by Eq.~(\ref{eq:metric}).
The dark matter and baryonic matter are assumed to interact only through gravity, and both components are treated as perfect fluids. Accordingly, the energy--momentum tensor is given by
\begin{align}
T^{\mu\nu} &= (\epsilon+p)u^\mu u^\nu + pg^{\mu\nu},\\
\epsilon &= \epsilon_{B}+\epsilon_{DM},\\
p &= p_{B}+p_{DM},
\end{align}
where the subscripts $B$ and $DM$ denote the baryonic and dark matter components, respectively. For the dark matter, we adopt the equation of state given by Eqs.~(\ref{eq: DMeos1}) and (\ref{eq: DMeos2}), while for the baryonic matter we employ the SLy equation of state. The equation for the metric function remains the same as Eq.~(\ref{eq: dnudr}) for pure dark stars. However, assuming that the two fluids are coupled only through gravity and each independently satisfies hydrostatic equilibrium, the Tolman-Oppenheimer-Volkoff equations become
\begin{equation}
\frac{dp_i}{dr} =  -\frac{(\epsilon_i+p_i)(m+4\pi pr^3)}{r(r-2m)},
\qquad
(i=B,DM),
\label{eq:dpidr}
\end{equation}
so that the pressure gradients of the two fluids evolve independently.
Here, The gravitational mass is given by $m(r)=\int_0^r \epsilon 4\pi r'^2 dr'$. On the other hand, the baryonic and dark matter masses are defined separately as
\begin{equation}
M_{DM}=m_{_{DM}}\int_0^R 4\pi r^2e^{\Psi}n_{_{DM}}dr
\label{eq: DMmass}
\end{equation}
\begin{equation}
M_{B}=m_{_B}\int_0^R 4\pi r^2e^{\Psi}n_{_B}dr
\label{eq: Bmass}
\end{equation}
In this work, we focus exclusively on the coupling between the scalar field and dark matter and assume that the scalar field does not couple to baryonic matter. Including a scalar-baryon coupling would obscure our primary objective of isolating the effects of the dark matter coupling. Moreover, such couplings are already subject to stringent constraints from pulsar observations. Consequently, in the third term of the effective potential, Eq.~(\ref{eq:Veff}), $- 4\pi \beta_d T e^{2\Phi}$, only the dark matter contribution to the trace of the energy-momentum tensor is included:
\begin{align}
T&\rightarrow T_{DM}\\\notag
&=-\epsilon_{_{DM}}+3p_{_{DM}}
\end{align}
\par
\begin{figure}
\centering
\includegraphics[width=\columnwidth]{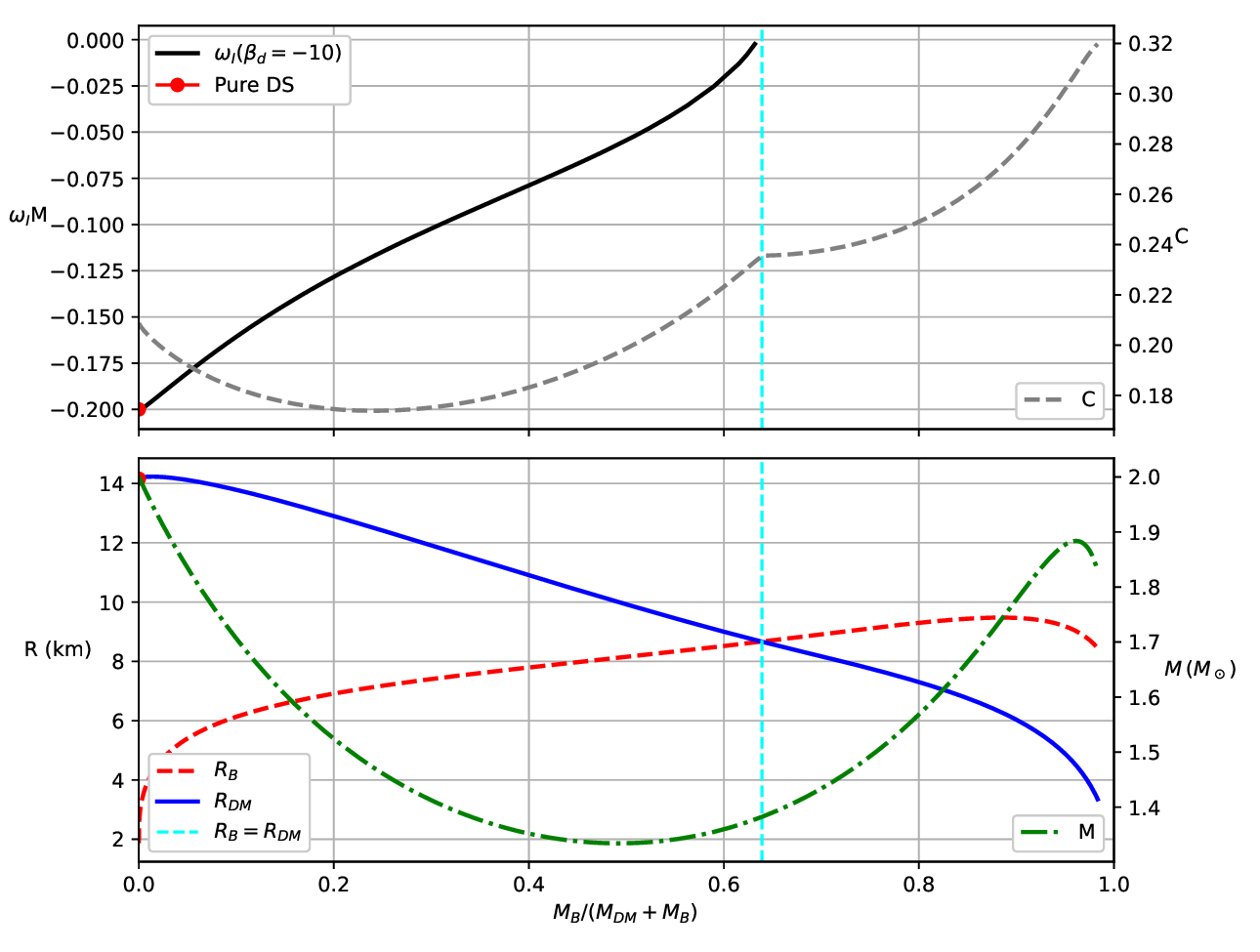}
\caption{\label{dnvc0.0066610632rhoc}Upper panel: Variation of $\omega_I$ (for $\beta_d=-10$) and the compactness $C$ as functions of the baryon mass fraction, $M_B/(M_B+M_{DM})$, for $\mu=4.1$ and $\kappa=9.2$, with $v_c$ held fixed. The left vertical axis denotes $\omega_I$, while the right vertical axis denotes the compactness. The red circle corresponds to the pure dark star with $(\mu,\kappa)=(4.1,9.2)$, and $M=2.0M_{\odot}$. Lower panel: The left vertical axis shows the radii of the dark matter sphere and the baryonic sphere, while the right vertical axis shows the stellar mass. As the baryon fraction increases, the envelope changes from dark matter to baryonic matter at $M_B/(M_B+M_{DM})\simeq0.64$.}
\end{figure}

Fig.~\ref{dnvc0.0066610632rhoc} shows the dependence of the imaginary part of the unstable mode, $\omega_I$, the compactness $C$, the gravitational mass $M$, the dark matter radius $R_{DM}$, and the baryonic radius $R_B$ on the baryon mass fraction, $M_B/(M_B+M_{DM})$, for DDSs with $(\mu,\kappa)=(4.1,9.2)$. The red circle corresponds to the pure dark star with $(\mu,\kappa)=(4.1,9.2)$, and $M=2.0M_{\odot}$. In this series, we fix the $v_c$ value at the center of this pure dark star and then add baryons. As the baryonic component is increased from this configuration, the dark matter radius $R_{DM}$ decreases, while the baryonic radius $R_B$ increases. The gravitational mass first decreases and then increases again. At $M_B/(M_B+M_{DM})\simeq0.64$, the baryonic radius becomes larger than the dark matter radius. The magnitude of the instability, $|\omega_I|$, is smaller for DDSs containing a larger baryonic fraction than for the corresponding pure dark star.

\begin{figure}
\centering
\includegraphics[width=\columnwidth]{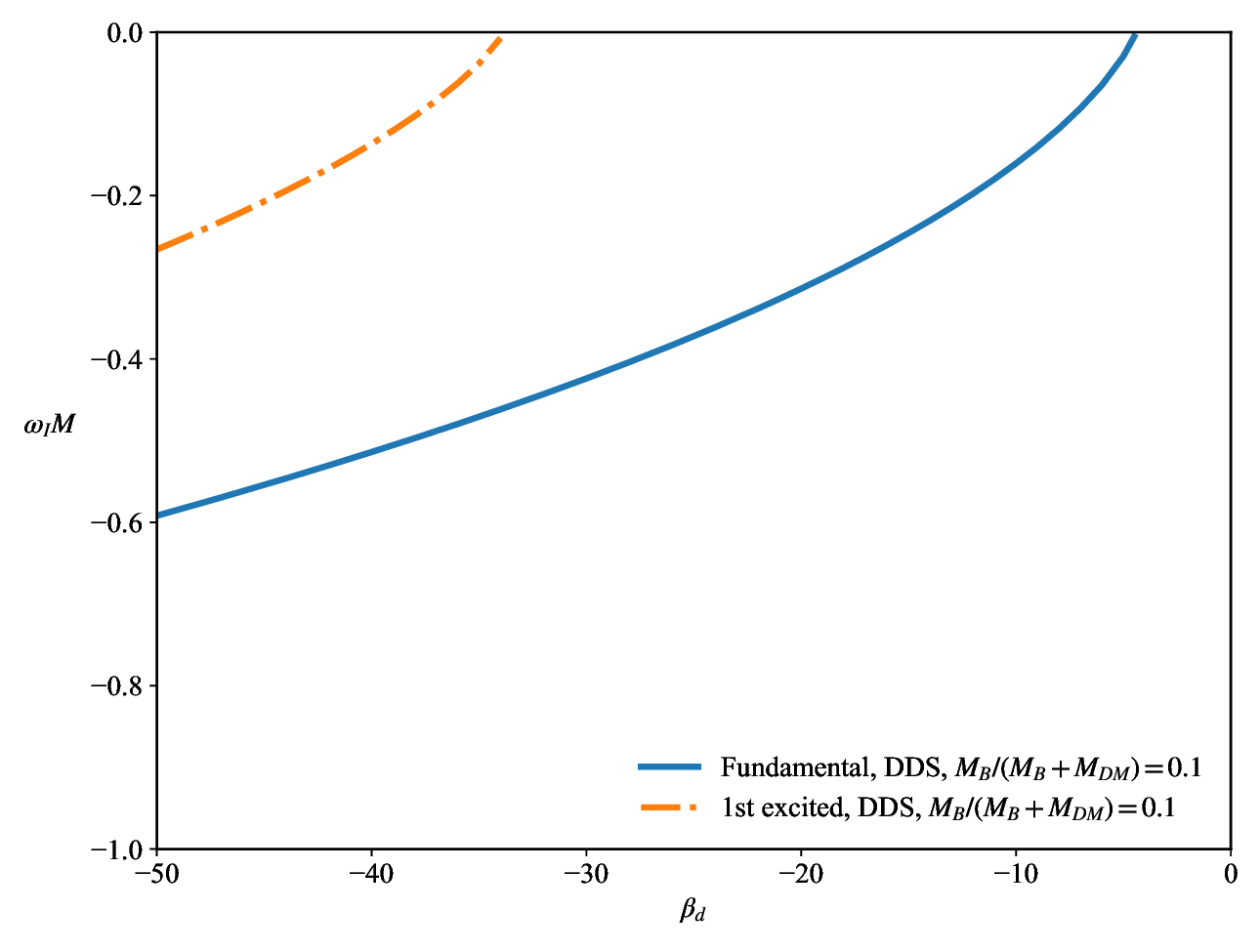}
\caption{\label{mb0.1omega}Variation of $\omega_I$ for a DDS. The stellar mass is $M=1.962\,M_{\odot}$, and the star contains a baryon mass fraction of $M_B/(M_B+M_{DM})\simeq0.1$. Dark matter particle parameters are $(\mu,\kappa)=(4.1,9.2)$.}
\end{figure}

\begin{figure}
\centering
\includegraphics[width=\columnwidth]{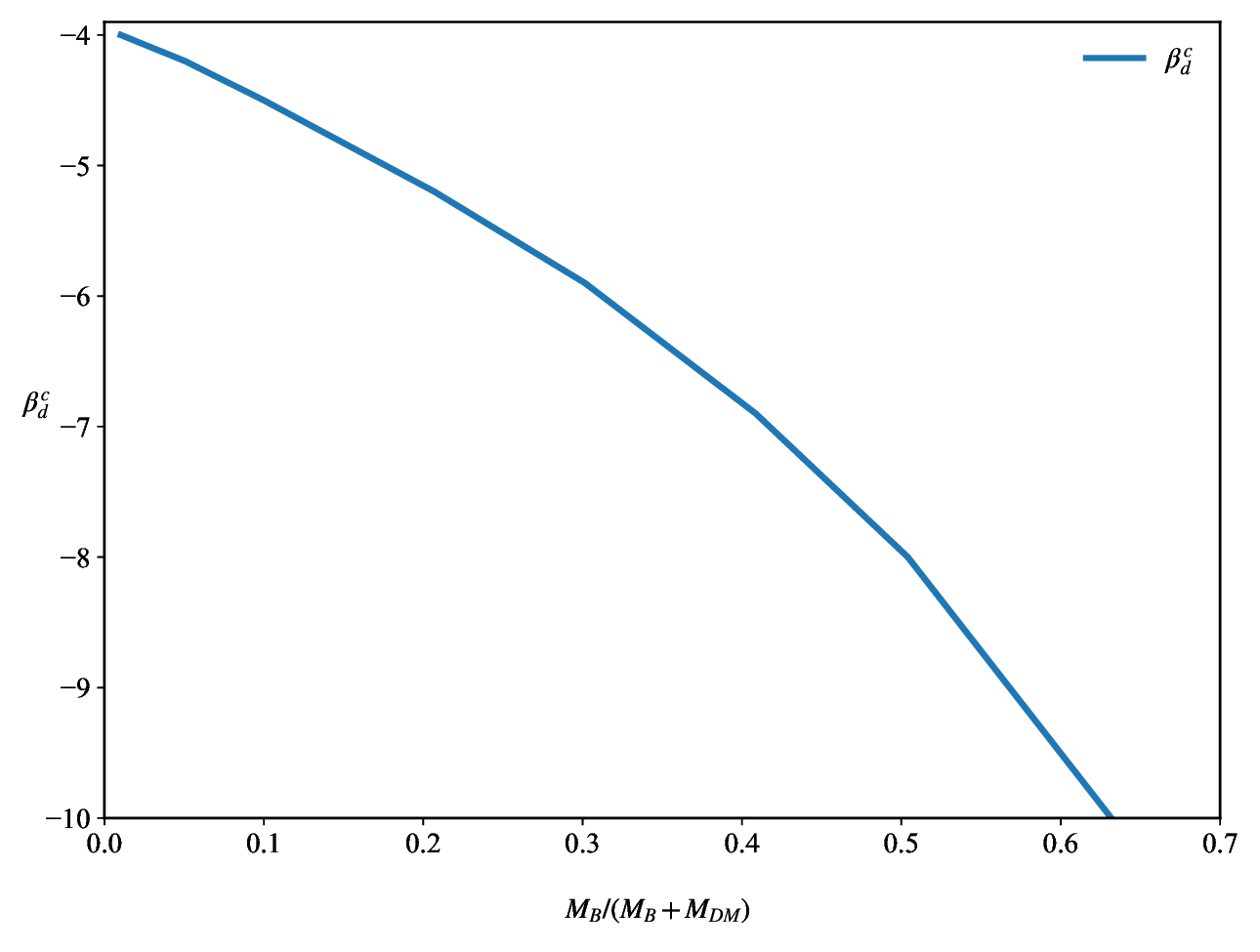}
\caption{\label{betac}Variation of the critical coupling parameter $\beta_d^c$ as a function of the baryon mass fraction, $M_B/(M_B+M_{DM})$. Dark matter particle parameters are $(\mu,\kappa)=(4.1,9.2)$.}
\end{figure}

Fig.~\ref{mb0.1omega} shows the dependence of $\omega_I$ on $\beta_d$ for a DDS with $M_B/(M_B+M_{DM})\simeq0.1$. Compared with the pure dark star shown in Fig.~\ref{M2m4.1k9.2omega}, the magnitude of $|\omega_I|$ is reduced over the entire range of $\beta_d$. Furthermore, for the pure dark star, the largest coupling parameter for which unstable modes exist is $\beta_d^c\simeq-4.0$. As the baryon fraction increases in DDSs, however, $\beta_d^c$ becomes more negative, as shown in Fig.~\ref{betac}. In other words, the tachyonic instability window becomes progressively narrower as baryonic matter is added to a pure dark star.

\begin{figure}
\centering
\includegraphics[width=\columnwidth]{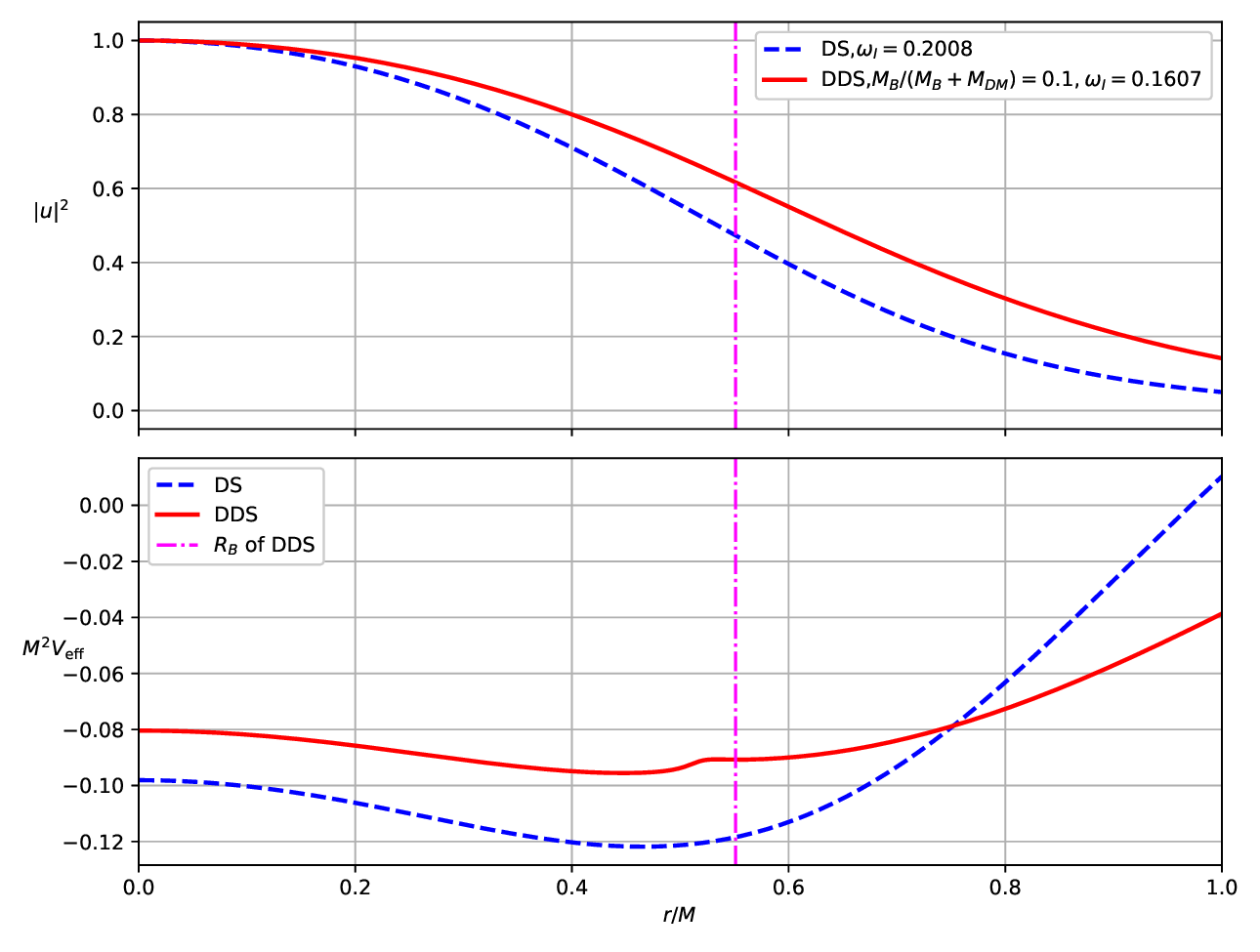}
\caption{\label{DS-DDS0.1}Upper panel: Comparison of the $|u|^2$ distributions of the fundamental mode at $\beta_d=-10$ for a pure dark star (DS) and a DDS with $M_B/(M_B+M_{DM})\simeq0.1$. Each distribution is normalized so that its maximum value is $|u|^2=1$. Lower panel: Corresponding effective potentials at $\beta_d=-10$. The magenta vertical dashed line indicates the baryonic surface. The stellar masses of the DS and DDS are $M=2.0\,M_{\odot}$ and $M=1.70\,M_{\odot}$, respectively. Dark matter particle parameters are $(\mu,\kappa)=(4.1,9.2)$.}
\end{figure}

\begin{figure}
\centering
\includegraphics[width=\columnwidth]{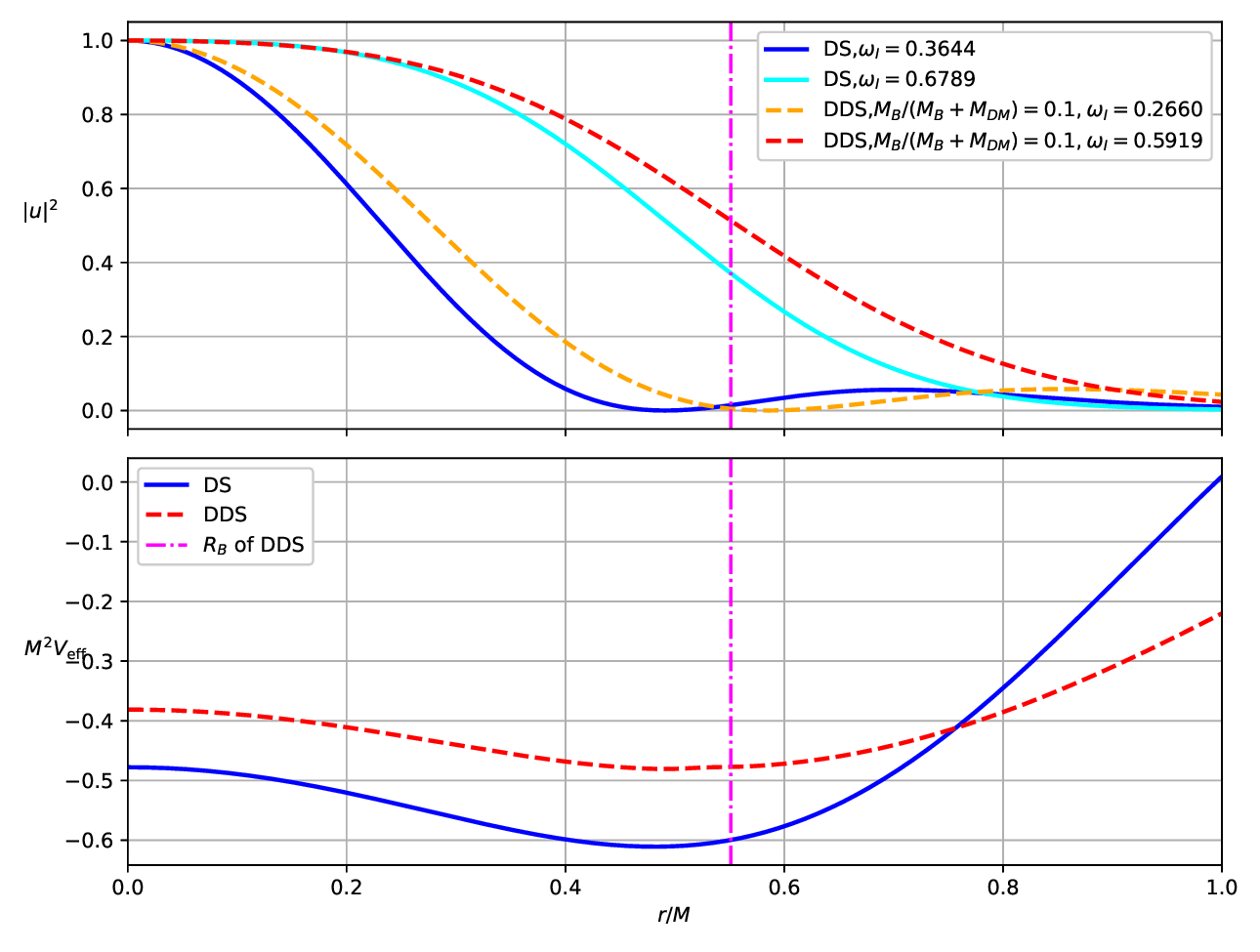}
\caption{\label{DS-DDS0.1-2}Upper panel: Comparison of the $|u|^2$ distributions of the fundamental mode and the first excited mode at $\beta_d=-50$ for a pure dark star (DS) and a DDS with $M_B/(M_B+M_{DM})\simeq0.1$. Each mode is normalized independently so that its maximum value is $|u|^2=1$. Therefore, the amplitudes of $|u|^2$ cannot be compared between different modes. Lower panel: Corresponding effective potentials at $\beta_d=-50$. The magenta vertical dashed line indicates the baryonic surface. The stellar masses of the DS and DDS are $M=2.0\,M_{\odot}$ and $M=1.70\,M_{\odot}$, respectively. Dark matter particle parameters are $(\mu,\kappa)=(4.1,9.2)$.}
\end{figure}

Figs~\ref{DS-DDS0.1} and \ref{DS-DDS0.1-2} compare the scalar-field distributions $|u|^2$ for a pure dark star and a DDS with $M_B/(M_B+M_{DM})\simeq0.1$ at $\beta_d=-10$ and $-50$, respectively. In both cases, the scalar-field distribution extends farther outward in the DDS than in the pure dark star. This behavior reflects the fact that the effective potential becomes shallower, making it more difficult for the unstable mode to remain localized inside the star.

\begin{figure}
\centering
\includegraphics[width=\columnwidth]{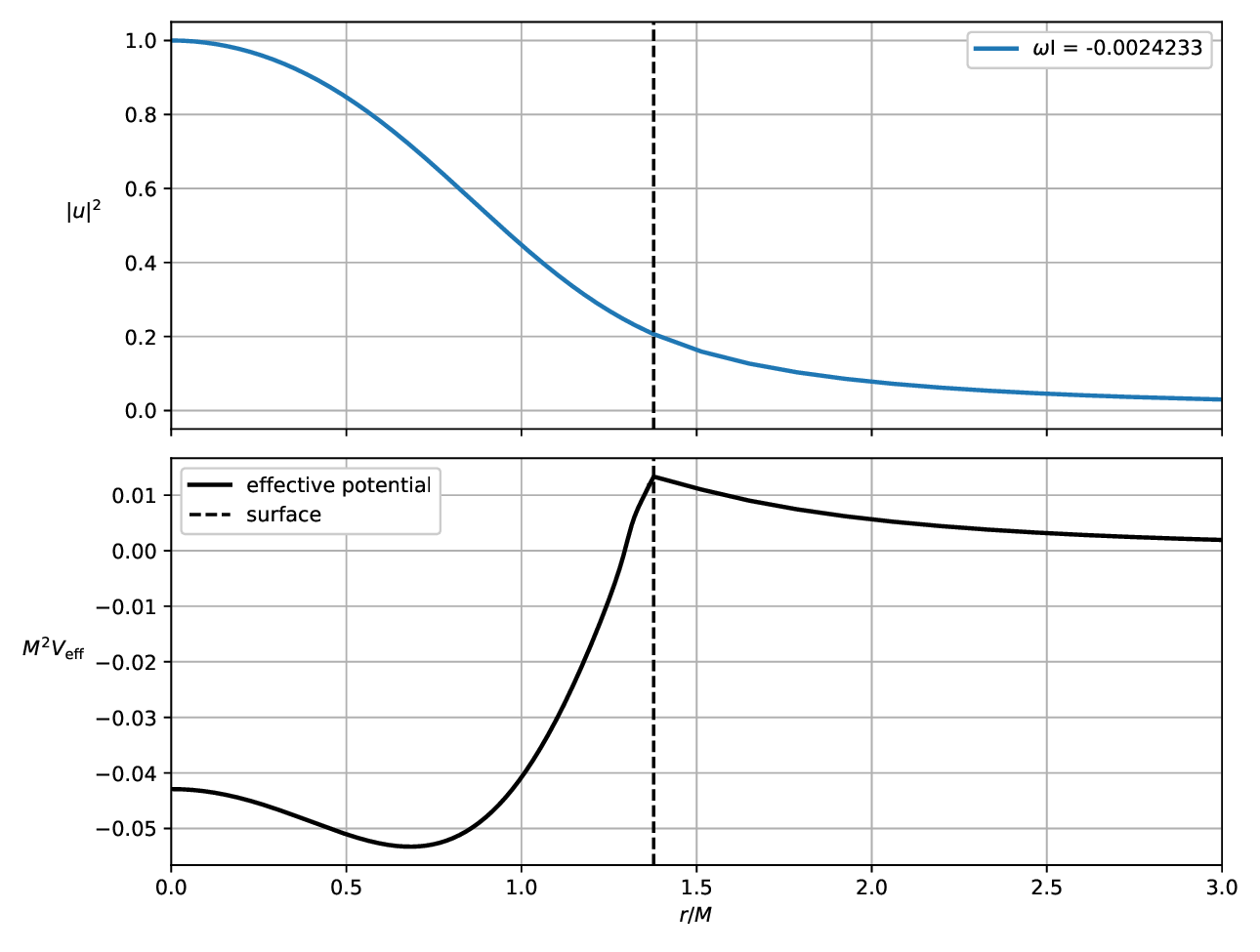}
\caption{\label{rdm=rbb-10psi}Upper panel: The distribution of $|u|^2$ corresponding to the mode with $\omega_I=-0.1281$ at $\beta_d=-10$. The distribution is normalized so that its maximum value is $|u|^2=1$. Lower panel: Effective potential at $\beta_d=-10$. The black vertical dashed line indicates the stellar surface. The stellar mass is $M=1.38\,M_{\odot}$, and the baryon mass fraction is $M_B/(M_B+M_{DM})\simeq0.64$. Dark matter particle parameters are $(\mu,\kappa)=(4.1,9.2)$.}
\end{figure}

As the baryon fraction is increased further, both $|\omega_I|$ and $\beta_d^c$ continue to decrease, and the excited mode disappears over the range $-50\le\beta_d\le0$. Both $|\omega_I|$ and $\beta_d^c$ attain their minimum values when $R_{DM}=R_B$. Fig.~\ref{rdm=rbb-10psi} shows the scalar-field distribution $|u|^2$ at $\beta_d=-10$ for this configuration. The distribution becomes substantially broader, indicating that the scalar field is no longer strongly localized but instead spreads over a much wider region.

Beyond the point where $R_{DM}=R_B$, the stellar configuration consists of a mixed dark matter--baryon core surrounded by a baryonic envelope, corresponding to the case in which dark matter accumulates at the center of a neutron star. In this regime, the DDS no longer exhibits a tachyonic instability. Therefore, in the model considered here, tachyonic instability occurs only in DDSs formed by adding baryonic matter to a dark star, whereas, neutron-star configurations with a dark-matter core do not exhibit the dark-matter-induced tachyonic instability.

\begin{figure}
\centering
\includegraphics[width=\columnwidth]{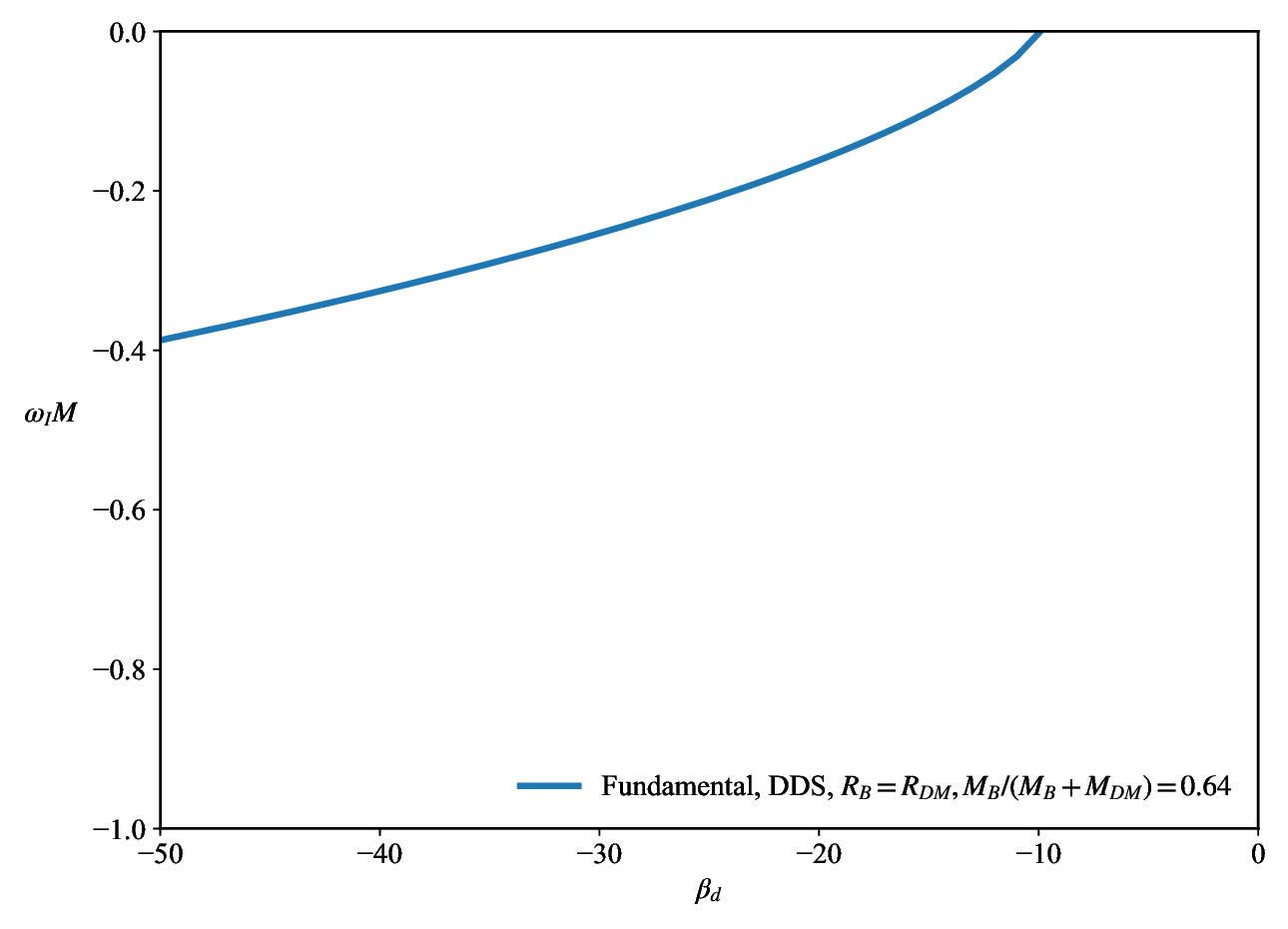}
\caption{\label{rdm=rdmomega}Variation of $\omega_I$ for a DDS. The stellar mass is $M=1.38\,M_{\odot}$, and the star contains a baryon mass fraction of $M_B/(M_B+M_{DM})\simeq0.64$. Dark matter particle parameters are $(\mu,\kappa)=(4.1,9.2)$.}
\end{figure}

\section{\label{con}Conclusion}
In this work, we investigated the tachyonic instability induced by the coupling between a scalar field and dark matter within the framework of scalar--tensor theories. We consider the linearized scalar perturbation around the GR background, neglecting the backreaction of the scalar perturbation on the metric and matter at first order. We consider both pure dark stars and mixed dark matter--baryon stars and found that both types of compact stars can exhibit tachyonic instability. For pure dark stars, the coupling parameter between the scalar field and dark matter, $\beta_d$, and the stellar mass, $M$, play crucial roles. Negative values of $\beta_d$ generate a negative potential well in the effective potential. This potential well triggers the growth of the tachyonic instability, and as $|\beta_d|$ increases, additional excited unstable modes appear in addition to the fundamental mode. On the other hand, the scalar-field distribution $|\Psi|^2$ and the growth rate of the instability are strongly influenced by the stellar mass through the shape and depth of the potential well. It is generally said that greater compactness tends to result in greater instability. However, our systematic survey over the equilibrium sequence shows that the instability growth rate is more strongly associated with the location of the equilibrium configuration along the mass sequence, becoming maximal near the maximum-mass stable configuration rather than at the largest compactness. Although various dark stars can be formed depending on the dark matter particle parameters $(\mu,\kappa)$, we found that, regardless of the combination of $(\mu,\kappa)$, the tachyonic instability reaches its maximum at the nearly maximum mass in the mass sequence. \par

We also compared pure dark stars with neutron stars. Although the two types of stars have similar masses, the effective potentials differ significantly in both shape and spatial profile. For both neutron stars and dark stars, the magnitude of $\omega_I$ varies with the stellar mass. The same applies to the threshold $\beta_d^c(\beta^c)$ at which the tachyonic instability occurs. Although both dark stars and neutron stars become more compact as their masses increase,he strongest instability appears. This suggests that these instabilities cannot be characterized by stellar compactness alone. Instead, they depend on the stellar configuration along each equilibrium sequence, which is determined by the underlying EOS. In other words, the pressure and energy density change due to the EOS, which in turn determines the energy-momentum tensor and, consequently, the effective potential, ultimately yielding $\omega_I$.\par

We also investigated whether the tachyonic instability can occur for $\beta_d>0$. We found that no unstable modes appear on the stable branch of dark stars when $\beta_d>0$. However, unstable modes can arise for positive values of $\beta_d$ on part of the unstable branch and in extremely stiff dark stars. In these configurations, the existence of unstable modes is determined by the central parameter $v_c$. As $v_c$ increases, the effective potential develops a potential well, allowing unstable modes to appear. Nevertheless, it is unclear whether a dark star can evolve to such configurations beyond the stable branch before undergoing gravitational collapse. Whether such dark stars can exist in nature and how they subsequently evolve under the tachyonic instability are beyond the scope of the present work and are therefore left for future study.\par

In astrophysical environments, it is necessary to consider DDSs in which a dark star contains a baryonic gas component. We found that the tachyonic instability becomes progressively weaker as the baryon fraction increases, and the critical coupling parameter $\beta_d^c$ shifts to more negative values compared with those of pure dark stars. In DDSs, the instability is primarily correlated with the baryon mass fraction along the DDS sequence considered here. Both $|\omega_I|$ and $\beta_d^c$ attain their minimum values when the dark matter and baryonic radii satisfy $R_{DM}=R_B$. Furthermore, once the baryonic radius exceeds the dark matter radius, corresponding to configurations in which dark matter is confined to the interior of a baryonic star such as a neutron star, the tachyonic instability is completely suppressed. Although baryons themselves do not have scalar coupling, because baryons alter the gravitational potential and the dark matter distribution, the potential well becomes shallower as a result, and the scalar field can no longer be trapped. This result indicates that neutron stars remain stable against the dark-matter-induced tachyonic instability considered in this work.\par

As noted above, there is no well-established theory that formalizes the relationship between scalar fields and dark matter. However, according to our research, even when assuming a weak interaction between the scalar field and dark matter, the tachyonic instability provides a necessary condition for the onset of spontaneous scalarization and may lead to nontrivial scalar configurations in the nonlinear regime. Our results only examine the conditions for linear onset of instability. Therefore, this study cannot determine what specific scalarized configuration the system will transition to, nor what the magnitude of the scalar charge will be. We assume that the background stellar structure remains that of GR because the scalar perturbation does not backreact on the metric or matter at linear order. However, if the scalar is more strongly coupled, the stellar structure itself would change from that in GR. Although phenomenological, such reports already exist \cite{2013PhRvD..88f3005F}, indicating that coupling with the scalar reduces the star's size itself. In such a situation, the stellar structure and the scalar-field configuration would influence each other and evolve self-consistently. Since this evolution is inherently nonlinear, it is considerably more difficult to analyze. This is a future work.

\bibliography{sods.bib}

\end{document}